\documentclass[aps,twocolumn,amsmath,amssymb,superscriptaddress,longbibliography]{revtex4-1}

\usepackage[dvips]{graphicx}
\usepackage[utf8]{inputenc}

\usepackage{soul,xcolor}
\setstcolor{red}

\usepackage{dcolumn}
\usepackage{physics}
\usepackage{bm}
\usepackage{graphics}
\usepackage{dirtytalk}
\usepackage{epsfig,color}
\usepackage[breaklinks,colorlinks,bookmarks=false,citecolor=red,linkcolor=blue,urlcolor=magenta]{hyperref}

\begin{document}
\title{Excited-Eigenstate  Entanglement Properties of  XX Spin Chains with Random Long-Range Interactions}
\author{Y. Mohdeb}
\email[]{y.mohdeb@jacobs-university.de}
\affiliation{Department of Physics and Earth Sciences, Jacobs University Bremen, Bremen 28759, Germany}  

\author{J. Vahedi}
\email[]{j.vahedi@jacobs-university.de}
\affiliation{Department of Physics and  Earth Sciences, Jacobs University Bremen, Bremen  28759, Germany}  
\affiliation{Department of Physics, Sari Branch, Islamic Azad University, Sari 48164-194, Iran}

\author{S. Kettemann}
\email[]{s.kettemann@jacobs-university.de}
\affiliation{Department of  Physics and Earth Sciences, Jacobs University
  Bremen, Bremen 28759, Germany}  
\affiliation{Division of Advanced Materials Science, Pohang University of Science and Technology (POSTECH), Pohang 790-784, South Korea}

\begin{abstract} 
Quantum information theoretical measures are useful tools for characterizing  quantum dynamical phases. However,  employing  them  to study  excited states of  random spin systems is a challenging problem.  
Here, we report results for  the entanglement entropy (EE) scaling of excited eigenstates of  random XX antiferromagnetic spin chains with long-range (LR) interactions decaying as a power law
with distance with  exponent $\alpha$.
To this end, we extend the  real-space renormalization group technique for excited states (RSRG-X) to 
 solve this problem with LR interaction. 
For comparison, we perform   numerical exact diagonalization (ED) calculations. From the  distribution of energy level spacings, as obtained by ED for up to $N\sim 18$ spins, we find indications of a delocalization transition at 
$\alpha_c \approx 1$ in the middle of the energy spectrum.
With RSRG-X and ED, we  show that for $\alpha>\alpha^*$ the entanglement entropy (EE) of excited eigenstates  retains a logarithmic divergence similar to the one observed for the ground state of the same model, while for $\alpha<\alpha^*$ EE displays an algebraic growth with the subsystem size $l$, $S_l\sim l^{\beta}$, with $0<\beta<1$. We find that $\alpha^* \approx 1$ coincides with the delocalization transition $\alpha_c$ in the middle of the many-body spectrum. 

An interpretation of these results based on the structure of the RG rules is proposed, which is due to  {\it rainbow} proliferation for very  long-range interactions $\alpha\ll 1$.   We  also investigate the effective temperature dependence of the EE allowing us to  study the half-chain entanglement entropy of eigenstates at different energy densities, where we find that the crossover
in EE  occurs at $\alpha^* < 1$.
\end{abstract} 

\maketitle

\section{Introduction} The magnetic properties of 
doped semiconductors as observed in magnetic resonance experiments\cite{Feher1955} motivated  P.W. Anderson to  address
the localization in interacting disordered systems, in particular  disordered interacting spin systems
\cite{anderson58}.  
 Fleishman and Anderson\cite{Fleishman1980}  
  showed that short-range interactions in an electron system with localized single-particle states might not destroy localization for some range of finite temperature $T$. 
In Refs. \cite{gornyi05,basko06}  it was argued  that
 many-body localization at finite temperature may result in a lack of thermalization.
Since then,  many-body localization (MBL) has become a flourishing research direction, for reviews see\cite{Abanin2019,Bhatt2021}. 

In a MBL phase, disorder, as modeled by randomness, can lead to localized states despite the presence of interactions. These states violate the eigenstate thermalization hypothesis (ETH)\cite{Deutsch1991} which states that the statistical properties
of physical observables of generic quantum Hamiltonians are the same as those predicted by the microcanonical ensemble. Thereby, given a subregion $A$ of $n$ spins in a chain, ETH requires that entanglement entropy $S_{A}$ scales with the volume of $A$, $S_A \sim n$\cite{Abanin2019}. Whereas in a MBL phase, excited eigenstates display an area law entanglement scaling $S_A\sim 1$. 

More recently, systems displaying a logarithmic divergence at finite energy density $S_A \sim \ln (n)$, and power law average correlations were discovered\cite{Vasseur2015,Vasseur2016}, they were dubbed {\it Quantum critical glasses} (QCG).
While, MBL is essentially established for some short-range models\cite{Luitz2015}, random bond spin chains with long-range interactions have to our knowledge not been investigated, although studies on long-range interacting spin chains with   random magnetic field have been done\cite{Schiffer2019,burin15,SafaviNaini2019}.

As finite temperature many-body localization is a property associated with excited many-body eigenstates, the entanglement properties of excited eigenstates are needed for random spin chains with long-range interactions.

To this end, we  introduce one of the most potent tools for studying one dimensional random Hamiltonians,  the strong disorder renormalization group (SDRG)\cite{dasguptama,monthus}. SDRG
has been widely used to study   properties of ground states of disordered spin chains with nearest neighbour interactions \cite{refael-entropy,fisher94} and beyond\cite{Yusuf2003,Lamas2006} and
more recently it was used to describe random spin chains with power-law decaying interactions \cite{ours,ourPRB,Mohdeb2020}.
This method  has recently been   extended to study 
 the whole set of eigenstates via the so-called RSRG-X (Real space renormalization group for excited states) procedure \cite{Pekker2014}.
 This technique is a powerful method to characterize the excited states of random interacting many-body systems, and can therefore be used to capture different dynamical phases of a given model at strong disorder. 
In particular, this procedure was previously used to study the entanglement properties of excited eigenstates of a random XX spin chain with nearest-neighbour interaction \cite{Huang2014,Pouranvari2015}, where it was found that the flow equations for the magnitude of the couplings are identical to the ground-state ones,
leading therefore, to a logarithmic divergence in the entanglement entropy. RSRG-X was also used to study the high energy states of the random bond XXZ spin chain\cite{Vasseur2016}.

Here, we consider a random bond-XX spin chain with couplings decaying with a power-law exponent $\alpha$.
Disorder is introduced in the model through a random choice of the positions of the spins on the
chain leading to randomness in the spin-spin couplings.
We study the excited eigenstate properties of  this model via both RSRG-X  and numerical exact diagonalization(ED). The ground state entanglement properties of such a system were previously studied in Ref. \cite{Mohdeb2020}  where it was found that  entanglement entropy at zero temperature displays a logarithmic enhancement irrespective of the values of $\alpha$. This was obtained via  an analytical formulation of SDRG, its numerical implementation and with  ED, respectively. 
In the next section, we are using numerical ED to  study the level spacing statistics as an indicator for MBL. In section \ref{RSRGX} 
 RSRG-X is introduced,  in section \ref{EE} it is applied to derive 
the  entanglement entropy  of this model 
and its dependence on subsystem size $l$. In section \ref{EEED}  the results  for the EE are presented as obtained with  exact diagonalization performed for up to $N\sim 14$ spins. With both methods, we find that  in the middle of the many-body spectrum, 
for any $\alpha>1$ the entanglement entropy diverges logarithmically and follows a Cardy law\cite{cardy04}  as observed similarly in the ground state\cite{ourPRB}. However, for $\alpha<1$ a sub-volume law for entanglement scaling is found  $S(l) \sim l^{\beta}$ with $0<\beta<1$.
Thus,  we find that the  crossover in entanglement entropy scaling 
at $\alpha ^{*} \approx 1$ coincides with  delocalization transition $\alpha \approx 1$
 in the middle of the many body spectrum. In section  \ref{EC} 
 results for the  entanglement  contour are presented and analyzed. In section  \ref{ST}  we  present an interpretation of the 
  strong power law violation of the area law for the entanglement entropy in terms of rainbow bond proliferation. In section  \ref{ETEE} we present results obtained at finite effective temperature, corresponding to an energy density away from the middle of the spectrum. In section  \ref{C}  we conclude. 

 We focus on  the  bond disordered XX-spin  chain with  long-range couplings,  defined by  the Hamiltonian
\begin{equation}\label{H}
H=\sum_{i<j}J_{ij}\left(S_{i}^{x}\,S_{j}^{x}+S_{i}^{y}\,S_{j}^{y}\right), 
\end{equation}
describing $N$ interacting $S=1/2$ spins that are placed randomly at positions ${\bf r}_i$ on a  lattice of length $L$ and lattice spacing  $a$, with density $n_0 = N/L = 1/l_0$, where $l_0$ is thus the average distance between  them. 
The couplings between all pairs of sites $i,j,$ are taken to be antiferromagnetic and long-ranged, decaying with a power law $\alpha$, 
\begin{equation} \label{jcutoff}
J_{ij} = J_0\left|({\bf r}_i-{\bf r}_j)/a\right |^{-\alpha}.
\end{equation}
We fix $J_0=1$ and $a=1$ in the following.
  
\section{Level spacing statistics}
\begin{figure}
\includegraphics[width=.5\textwidth]{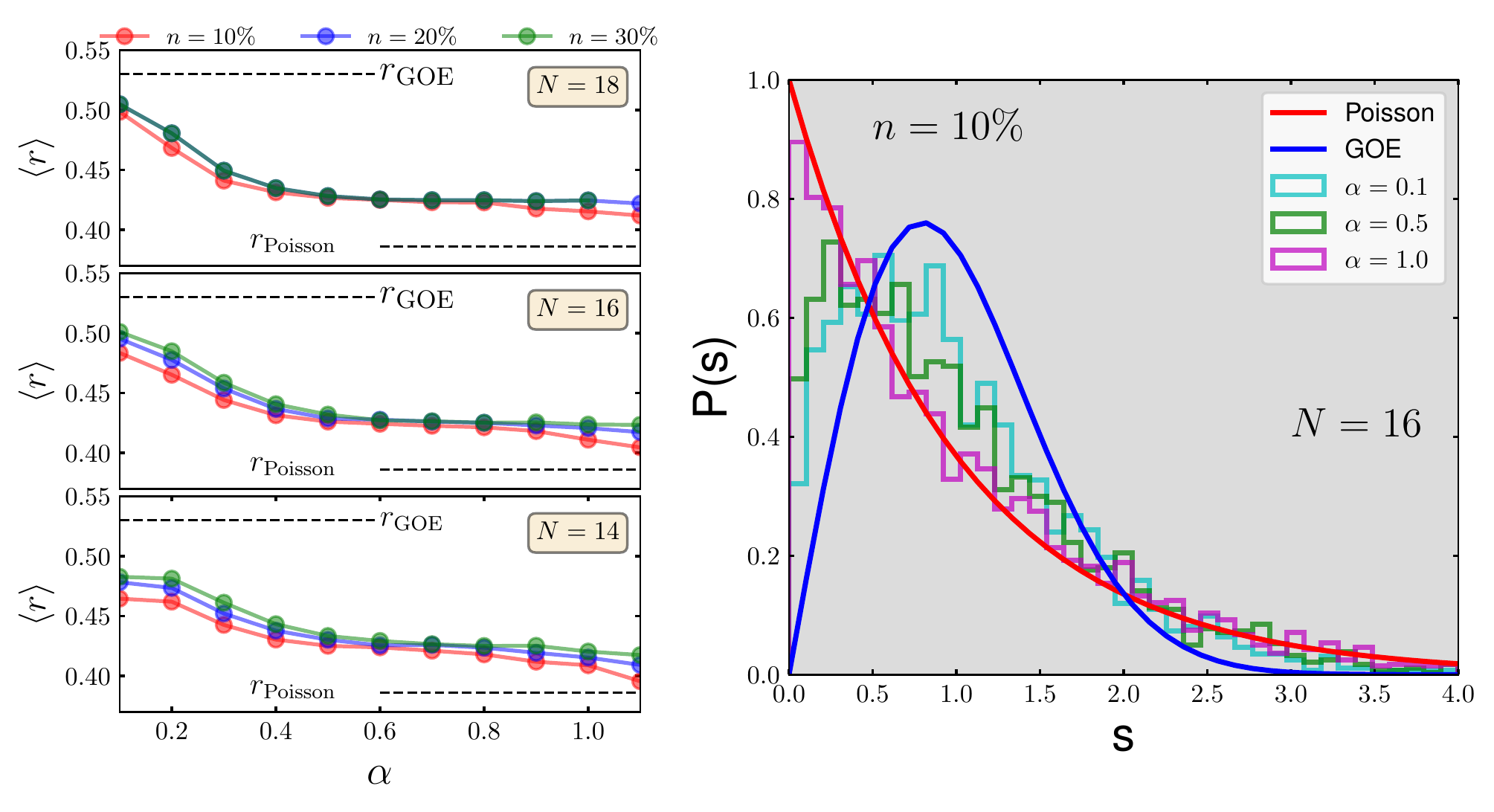}
\caption {Left: Adjacent gap ratio for various values of $\alpha$ and filling factors. Right: Probability distribution $P(s)$ of level spacings   $s=E_{n+1}-E_n$  between consecutive unfolded eigenvalues~\cite{Gubin2012}. Note, particularly that for the probability distribution, we  diagonalize the Hamiltonian Eq. (\ref{H}) in  the  $S_{{\rm tot}=0}$-subspaces, as  the projection of the total spin along the $z$ axis, $S_{{\rm tot}}=\sum_i S^z_i$, is conserved.  The results  are averaged over $1000$ disorder realizations. For the adjacent gap ratio $N_s=50$ states are taken in the middle of the many-body spectrum.}
\label{LSS}
\end{figure}

Level spacing statistics is known to be a convenient indicator to identify whether a system is in a delocalized phase, a  localized phase, or another regime. 
Disordered Hamiltonians  with  time-reversal and spin symmetry are known to be described
 by the   Gaussian orthogonal ensemble (GOE),  without time reversal symmetry by the Gaussian unitary ensemble (GUE) level spacing statistics, when they are in the ergodic regime as characterised by energy level repulsion. In a localized phase, the level spacing statistics obeys  rather the Poisson distribution, indicating the absence of level repulsion\cite{levels,levels2}. Via exact diagonalization, we compute the distribution $P(E_{n+1}- E_n )$
of the energy level spacings of the model Eq. (\ref{H}) for different filling factors $n_0=\frac{N}{L}$ and various values of the exponent $\alpha$. Further, we calculate the ratio of consecutive gaps of distinct energy levels, also known as the adjacent gap ratio: 
\begin{equation}
    r = \frac{1}{N_s} \sum_n \frac{min(E_{n+1}-E_n, E_n-E_{n-1})}{max(E_{n+1}-E_n, E_n-E_{n-1})},
\end{equation}
where $N_s$ is the number of states in the spectrum. The value of $r$, averaged over several disorder realizations, is known to be $0.5307$\cite{Atas2013} for the GOE and around $0.386$ for the Poisson distribution. Results are shown in Fig. \ref{LSS}, together with both limiting values. We see a crossover between a regime where the level spacing approaches the GOE for sufficiently small values of $\alpha$ and a Poissonian regime for $\alpha>1$. In a large interval of $\alpha<1$ it  is in an intermediate regime. These results suggest the possibility of a delocalization-localization transition occurring at $\alpha_c\approx 1$  in the middle of the  many-body spectrum.

\section{RSRG-X method in the Presence of Power Law Couplings} \label{RSRGX}
Next,  we intend to evaluate the entanglement entropy (EE) scaling with subsystem sizes as a function of the exponent $\alpha$. EE provides a quantitative tool to characterize how information is spread from one part of the system to
another.
To this end, we  use both a renormalization group scheme and numerical exact diagonalization.
We first describe how to  apply the RSRG-X to this model with antiferromagnetic long range interactions. For the ground state, at each step of SDRG\cite{monthus,hoyos07,Mohdeb2020}, we identify the largest  coupling $J_{ij}=\Omega$ and put the  two  spins that are coupled by $\Omega$ in the lowest energy state,  a singlet state. Assuming  strong disorder, the coupling of these two spins is much larger than  the remaining couplings. We  therefore  treat them as perturbations  and derive an effective Hamiltonian for the  remaining spins.
This procedure is continued until we form $N/2$ singlet pairs. Thereby the ground state of the system is approximated as the tensor product of these singlet states within SDRG. 
\newline
Excited eigenstates can  be obtained by a modified
version of SDRG, known as RSRG-X. In this
method, the two spins with the largest (in magnitude) coupling constant are chosen to be in one of their four  eigenstates (one singlet, three triplet states) with energy $E$ by the Boltzmann distribution, depending on an effective temperature $T$, so that  for each  of the singlet and triplet states, there is a corresponding probability associated with the parameter  $T$. The effective couplings for the other spins depend then
on the choice of the  state  for the two spins.
To determine the RG rules for excited states, we make use of   degenerate perturbation theory. Namely, a Schrieffer-Wolf transformation (SWT)\cite{Bravyi2011} is applied. SWT is a perturbative unitary rotation that eliminates 
off-diagonal elements of the Hamiltonian ${H}$ with respect to a {\it strong} piece $H_0$. 
Specifically, one writes $H = H_0+V$, where $V$
is parametrically smaller than $H_0$, as denoted by  $V\in \mathcal{O}{}{\lambda}$.  Then, one searches for a unitary operator $e^{i S}$
such that $[e^{iS} H e^{-iS}, H_0] = 0$ to
the desired order in $V$. This results in  a self-consistent equation for $S$ at each order, which can then be   solved. Further, one  projects onto an eigenstate subspace of $H_0$, and finds an effective Hamiltonian for the remaining degrees of freedom.
Projecting onto the lowest-energy eigenstate at each step of RSRG-X gives the ground state of the model reproducing the usual SDRG scheme, whereas projecting onto other subspaces allows access to generic excited eigenstates.

\begin{figure}
  \includegraphics[width=0.5\textwidth]{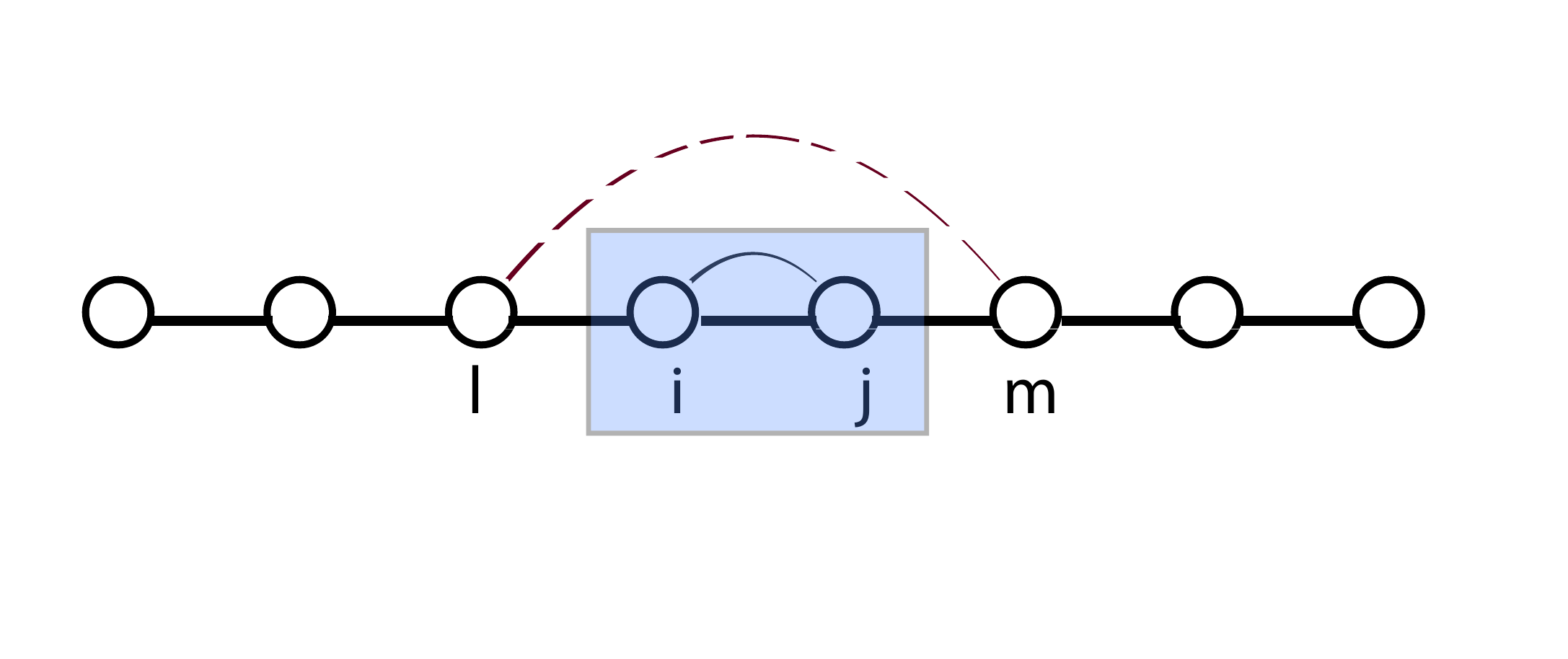}   \vspace{-1.5cm}
\caption{Decimation  of the strongest-coupled 
 pair $i,j$ (highlighted in blue)  generates effective couplings between other spins  $l,m$ (brown dotted line). }
  \vspace{-0.6cm}
\label{rgnnl}
\end{figure}

Here, we give an overview of SWT, and derive the RSRG-X rules for the Hamiltonian Eq. (\ref{H}).
We seek for an operator $S$, such that $[e^{iS} H e^{-iS}, H_0] = 0$. Expansion of the unitary rotation of $H$ in  $S$ gives:
 \begin{equation}\label{Hausdc}
\begin{aligned}
  &  e^{iS}H e^{-iS}= H_0+V+  [iS,(H_0+V)] \\ &
   -\frac{1}{2}\{S^2,H_0+V\}+ S (H_0+V) S +...
     \end{aligned},
 \end{equation}
where $\{ ., .\}$ stands for the anticommutator.
We now expand S in powers of  $\lambda$, 
$S=\sum_n S_{(n)}$ with $S_{(n)}\in \mathcal{O}{}{\lambda^n}$.
The condition $[e^{iS}H e^{-iS},H_0]=0$ then fixes the form of $S_{n}$.
At first order, we find
\begin{equation}
S_{(1)}=i \sum_{\alpha\neq\beta}\frac{|\alpha\rangle \langle\alpha|V |\beta\rangle \langle\beta| }{E_{\alpha}-E_{\beta}},
    \label{SW}
\end{equation}
where $\alpha$, $\beta$ are eigenstates of $H_0$, the  singlet or one of the triplet states,
formed by the spins with the largest coupling, and $E_{\alpha}$, $E_{\beta}$ are their corresponding eigenvalues.
The effective Hamiltonian is then given by the first order expansion of Eq. (\ref{Hausdc}), namely
\begin{equation}
H_ {eff}=e^{iS_{1}}H e^{-iS_{1}}-H_0= 
\sum_{\alpha}|\alpha\rangle \langle\alpha| (V+ [iS_1,V])|\alpha\rangle \langle\alpha|.
\end{equation}
Having described the framework for performing our degenerate perturbation theory, we now apply this procedure to the Hamiltonian in Eq. (\ref{H}).
In the case of the LR spin chain, one obtains the first order effective Hamiltonian and therefore deduces the  RG rules  for the different choices of projections for $H_0$. Crucially, we note that the effective
Hamiltonian keeps  the  XX-form of the original one, allowing  the procedure to be readily iterated.
The results are summarized below. Here, $(i,j)$ denotes the pair with the strongest coupling, as shown in Fig. \ref{rgnnl}.
If  the pair $ (i,j)   $ is projected onto $|\uparrow\uparrow\rangle$ or $ |\downarrow\downarrow\rangle$ we obtain
\begin{equation}
\label{upup}
    (J_{lm})'= J_{lm}- \frac{J_{i l} J_{jm} +J_{im} J_{jl} }{J_{ij}}.
\end{equation}

If $ (i,j) $ is in $\frac{1}{\sqrt{2}}(|\uparrow\downarrow\rangle+ |\downarrow\uparrow\rangle)$
\begin{equation}
\label{updown}
    (J_{lm})' =   J_{lm} + \frac{(J_{li}+J_{lj})(J_{im}+J_{jm})}{J_{ij}}.
\end{equation}
If $ (i,j) $ is in $\frac{1}{\sqrt{2}}(|\uparrow\downarrow\rangle- |\downarrow\uparrow\rangle$), we recover the result of Ref. \cite{ours,Mohdeb2020}
\begin{equation} \label{jeff}
               (J_{lm})' =   J_{lm} - \frac{(J_{li}-J_{lj})(J_{im}-J_{jm})}{J_{ij}}.
              \end{equation}

The RSRG-X procedure consists in  reiterating this scheme, and updating the value of all  couplings at each RG step. Thereby a generic RSRG-X eigenstate is obtained by taking the tensor product of all singlet  and triplet pairs obtained along the RSRG-X flow. We note that, at first, due to the   RG rule  in Eq. (\ref{upup}) the  generation of  ferromagnetic couplings along the  RG flow is possible. This is also the case for the RSRG-X applied to the nearest neighbour XX spin chain\cite{Huang2014}.
However,  as the decimated pairs are the ones with the largest energy gap in their spectrum, this 
turns out not to be relevant. 

\section{ Entanglement entropy} \label{EE}
The entanglement entropy of a pure state $\rho_{AB}$ is defined as $S(\rho_{A})=-Tr(\rho_{A} \ln(\rho_A))$, where
$\rho_A = Tr_{B}(|\psi\rangle\langle\psi|)$ is the reduced density matrix after tracing out a part $B$ of the system and $|\psi\rangle$ is the considered eigenstate.  It is an important diagnostic to identify phase transitions in disordered quantum systems\cite{Vasseur2015,Schiffer2019}.
We aim to evaluate the average entanglement entropy
of a part $A$ of length $l$ of a spin chain, 
$\langle S_{l}(T)\rangle$ with eigenstates $|\psi_i\rangle$ sampled from the Boltzmann distribution at effective temperature $T$\cite{Huang2014},
\begin{equation}
   \langle S_l(T) \rangle = \langle \sum_{i}\frac{exp (- \langle\psi_i| H/T|\psi_i\rangle)S_l(|\psi_i\rangle)}{Z(T)} \rangle,
\end{equation}
where $\langle . \rangle$ stands for the disorder average.
For this purpose, we use the fact that
if the spins $i$ and $j$ are projected onto a singlet state or to the entangled
triplet state $\frac{1}{\sqrt{2}}(|\uparrow\downarrow\rangle+ |\downarrow\uparrow\rangle)$, a unit of entanglement is generated, whereas the two triplet states  $|\uparrow\uparrow\rangle,|\downarrow\downarrow\rangle$ do not contribute to the entanglement.
At a given parameter $T$, we sample the 
states from  the Boltzmann distribution for every disorder realization
by choosing at each step of the RSRG-X procedure 
to project the most  strongly coupled pair of spins to a singlet or one of the triplet states according to the respective largest probability determined by their energy $E$ and parameter  $T$. 
For each disorder realization, we  then take the average of the generated entanglement for different eigenstates at fixed $T$ .
Note that $\langle S_l(T) \rangle$ is not the entanglement entropy of the thermal mixed states, but rather the EE of pure eigenstates sampled around a certain  energy region as determined by the effective  "temperature" $T$.

\begin{figure}[ht]
\centering
\includegraphics[width=.5\textwidth]{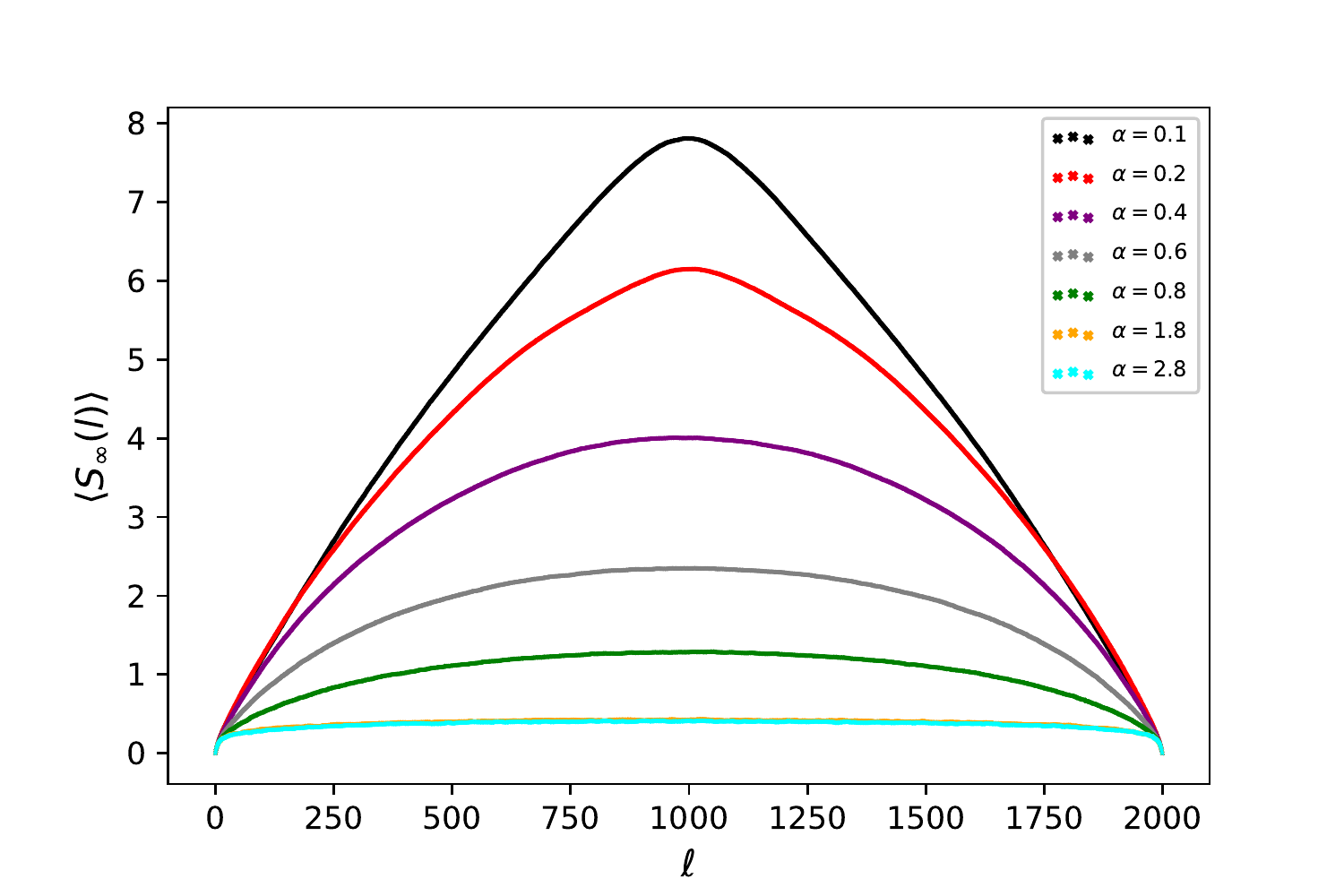}
\caption{Average entanglement entropy of excited eigenstates at infinite effective temperature  as a function of the partition length $l$ (physical distance), obtained from numerical RSRG-X for the long-ranged XX-chain with open boundary conditions  for $N=200$ spins
for various values of $\alpha$. The filling factor was fixed to  $N/L=0.1$. The average was evaluated over $ 10000$ disorder realizations, and $100$ sampled states for each disorder realization.   }
\label{realEEvdens}
\end{figure}

The results are shown in Fig. \ref{realEEvdens}.
Here we are working in the limit of infinite effective temperature $T \rightarrow \infty,$
where all eigenstates are sampled equiprobably. Thus, this probes the middle of the energy spectrum\cite{Vasseur2015}.
The average entanglement entropy of eigenstates  as a function of the partition length $l$ (physical distance) is displayed
for different values of $\alpha$. Here, the system is divided into two
subsystems of sizes $l$ and $L-l$ respectively and EE is computed for $N=200$ spins. A clear increase of EE is observed as $\alpha$ is lowered for $\alpha<1$, while 
 the results  for $\alpha= 1.8$ and $\alpha=2.8$ are almost identical. 

\begin{figure}
\centering
\includegraphics[width=.45\textwidth]{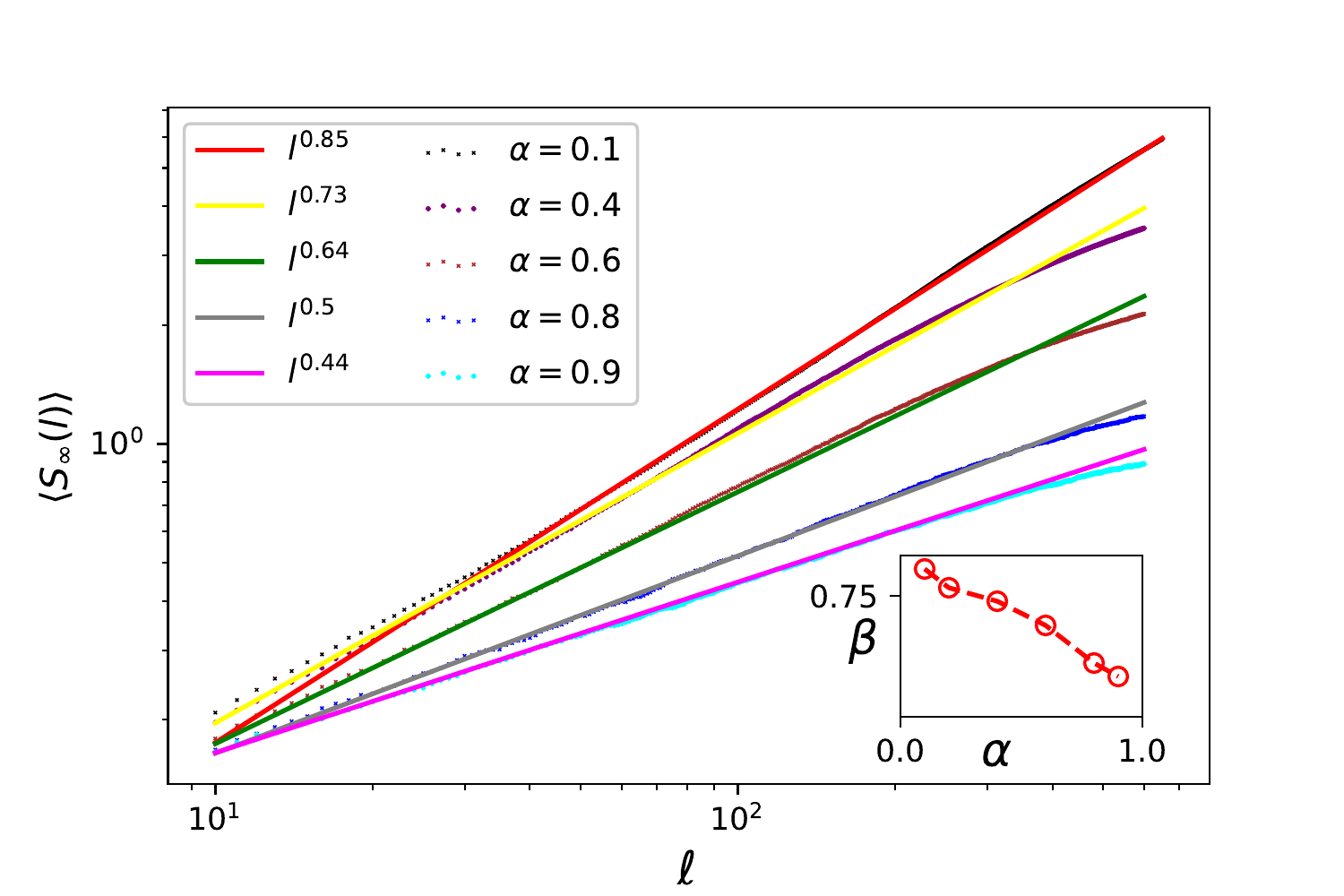}
\includegraphics[width=.45\textwidth]{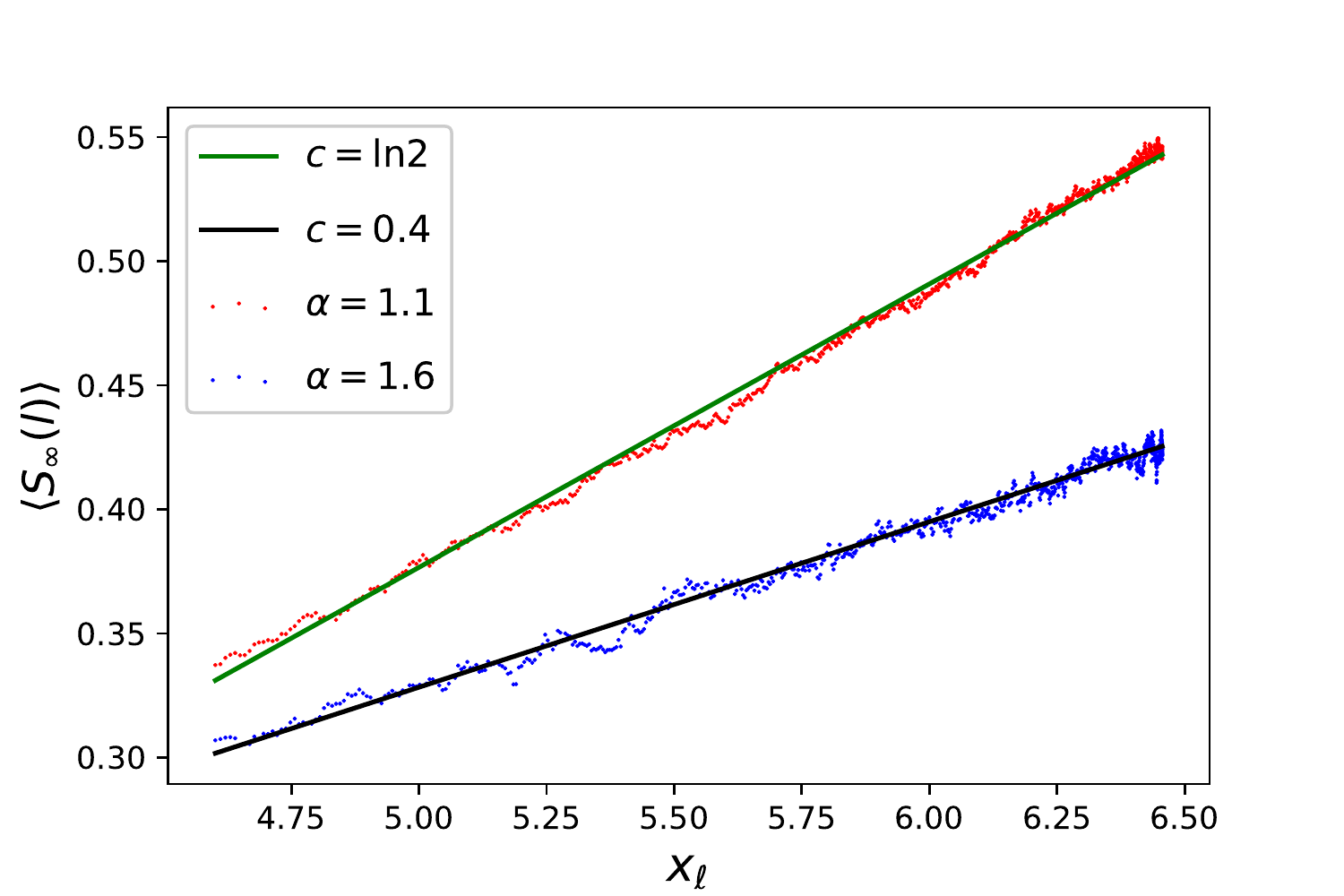}
 \caption {Top : Fig. \ref{realEEvdens} on log-log scale for $\alpha<1$, the inset plot shows $\beta$ the exponent in $S \sim l^{\beta}$ as a function of $\alpha$. Bottom: EE as a function of the logarithm of the chord distance $x_l$ for $\alpha=1.1$ and $\alpha=1.6$. $c$ stands for the effective central charges.}
\label{PowerEE}
\end{figure}

For $\alpha < 1$  the EE grows with $l$ slower than 
volume law, which would be linear in $l$.
We observe  in Fig. \ref{PowerEE} (top) 
that on a log-log scale the EE for each $\alpha< 1$ can be fitted with a straight line, $\ln(S_l ) = \beta \ln(l) +a$. Thus, the EE obeys a power-law dependence on partition length, $S_l= a l^{\beta}$,
with   fitted   power smaller than one,  $\beta <1$, for all $\alpha <1$.   Such a power-law behavior is due to the presence of sufficiently long-range interactions in this model, and was previously observed for the  ground states of the power-law random banded model (PRBM)\cite{Mirlin2000} as reported in  Ref. \onlinecite{PouranvariIPRB}, but also for the ground states  of free fermions with  long-range hoppings and  XX spin chains with long-range couplings  of random sign and amplitude  \cite{Roy2019,Gori2015}.

For $\alpha>1$ the EE  displays a logarithmic behavior as in the nearest neighbour case \cite{Huang2014}. To confirm this, we  plot the average entanglement entropy for $\alpha=1.1$ and $\alpha=1.6$ in Fig. \ref{PowerEE} (bottom) as a function of the logarithm of the chord distance, $x_l=\ln(\frac{L}{\pi} sin(\frac{\pi l }{L}))$.
 Indeed, the entanglement entropy turns out to be linear on this scale with a slope $a=\frac{c_{eff}}{6}$ corresponding to a Cardy law  for  open boundary conditions,\cite{cardy04} with an effective central charge $c_{eff}=\ln 2$ for $\alpha=1.1$ and $c_{eff}=0.4$ for $\alpha=1.6$, as is plotted as dashed lines. 

 The exponent of the power law  EE divergence, $\beta$, is found to  decrease  with increasing
  $\alpha$  for $0<\alpha<1$ as can be seen in the inset plot of Fig. \ref{PowerEE}.
  For $\alpha >1$
  the  EE scaling turns into a logarithmic  dependence on $l$ for any $\alpha>1$, as shown exemplary in Fig. \ref{PowerEE} (bottom) for $\alpha=1.1$ and $\alpha=1.6$.
We mention that the results are presented  here 
for the entanglement entropy as a function of the real distance $l = r_i-r_j$, where $r_i$ is the position of spin $i$, to be contrasted with the index distances $n = |i -j|$ between the spins. Similar scaling is observed as a function of the index distances.

\section{ED study of entanglement entropy scaling} \label{EEED}
\begin{figure}
\centering
\includegraphics[width=.5\textwidth]{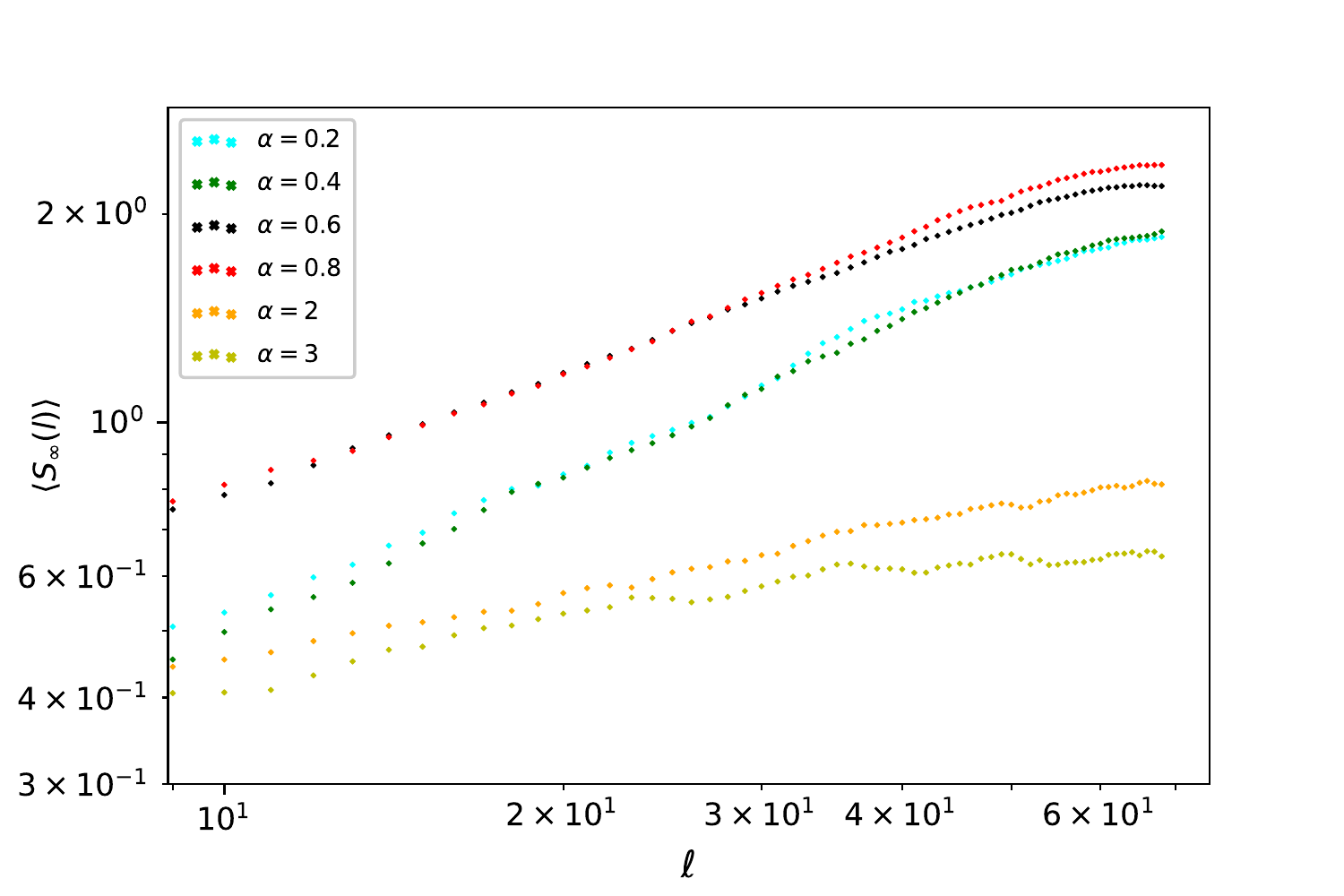}
\caption{ED average entanglement entropy in the middle of the energy spectrum  for $L=140$ sites and a filling factor $N/L=0.1$ for various values of $\alpha$. The results are obtained for $250$ realizations, averaging over $20$ states for each disorder point.}
\label{ee}
\end{figure}

\begin{figure}
\hspace*{-1cm}
\includegraphics[width=.55\textwidth]{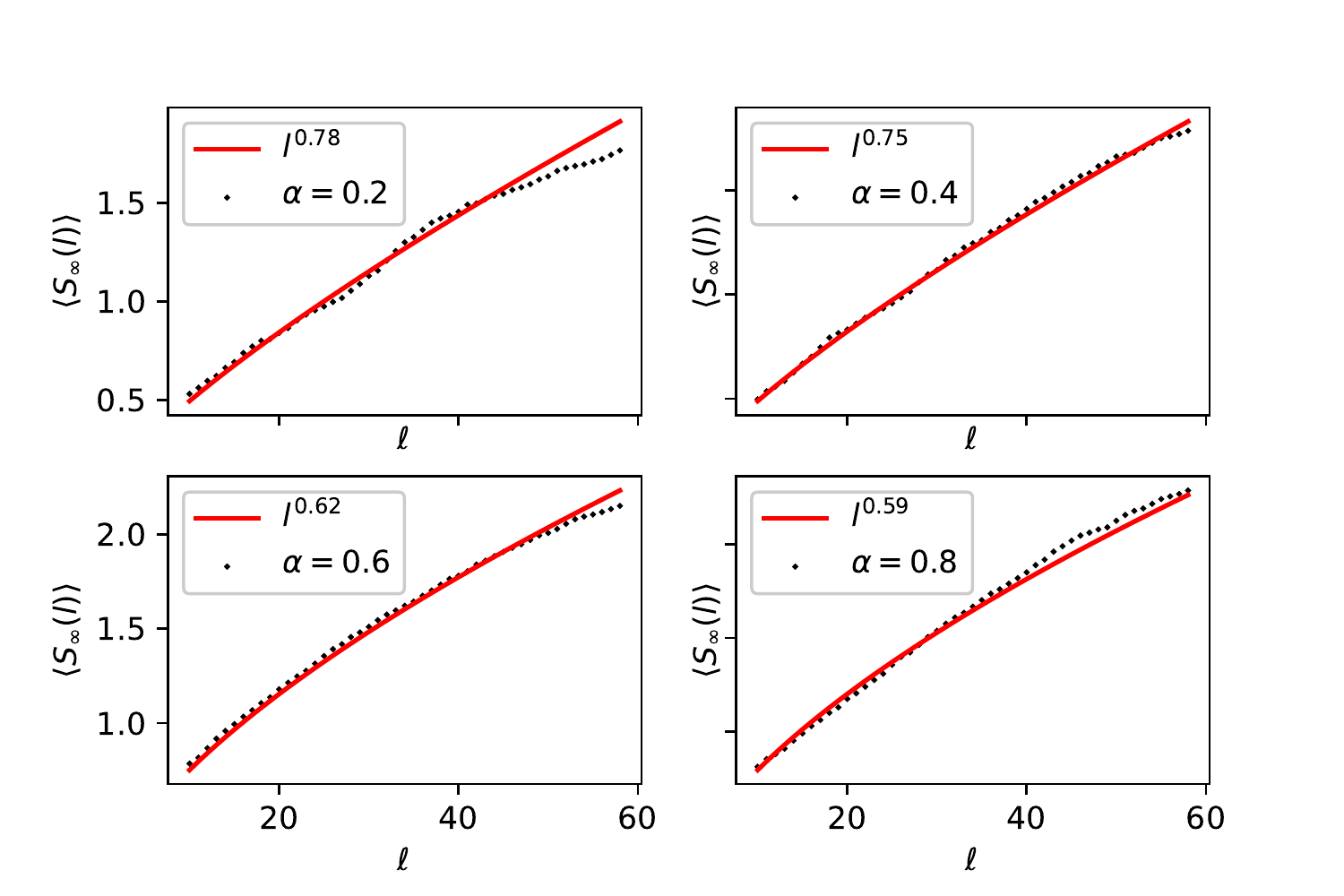}
\caption{ EE on log-log scale for $\alpha<1$ as obtained by numerical ED. The red curves are functions of the form $a l^{\beta}$ with fitted $\beta$ as indicated.}
\label{EDsa}
\end{figure}

\begin{figure}
\includegraphics[width=.4\textwidth]{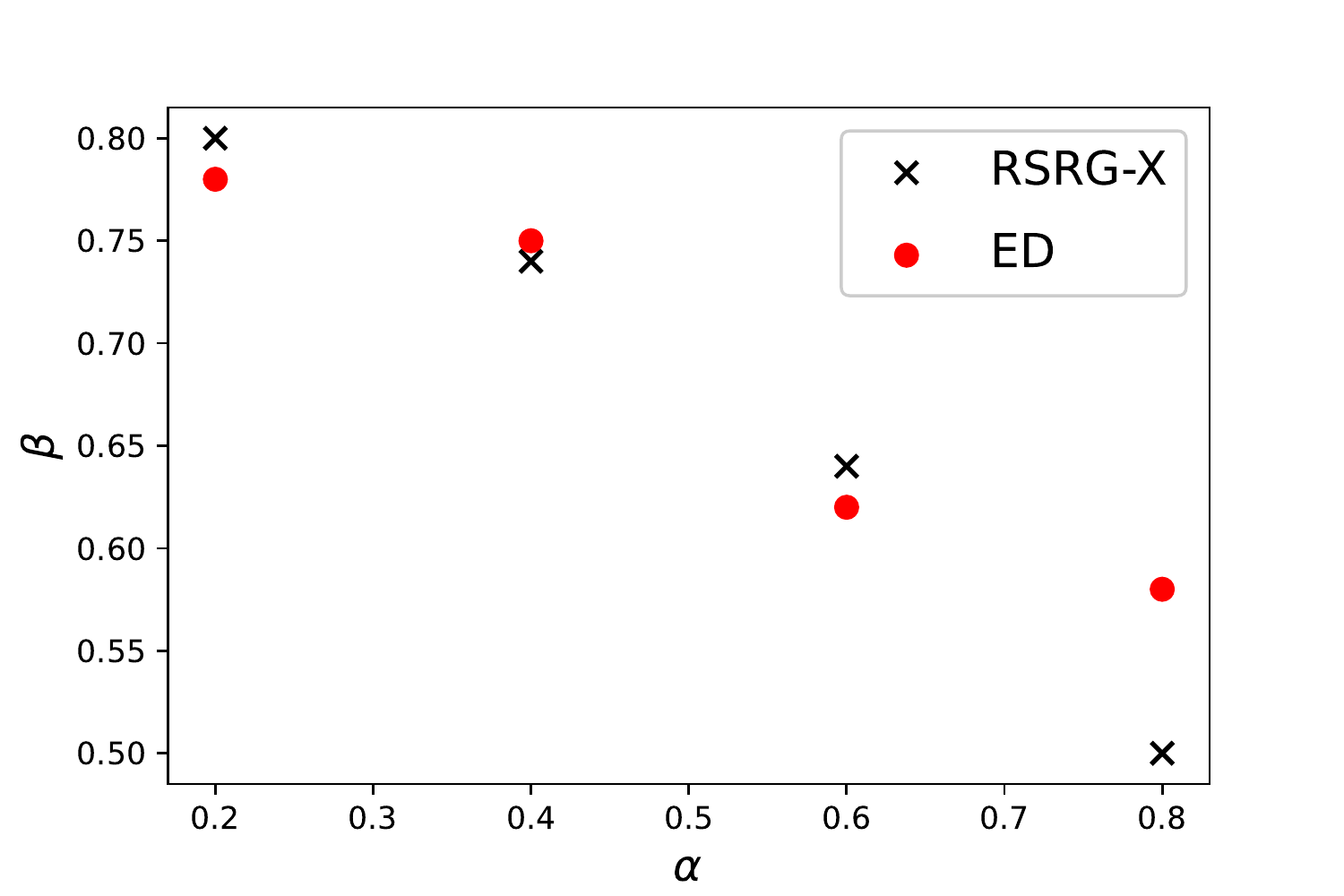}
\caption{ The exponent $\beta$  as function of $\alpha<1$, 
as obtained by fitting the ED and RSRG-X results to  the  function
$S(l)\sim l^{\beta}$.}
\label{EDRGcom}
\end{figure}

Although limited to small system sizes, numerical exact diagonalization has been extensively used   to study the properties of  excited states  of random and disordered spin chains\cite{Oganesyan2007,pal10,Schiffer2019,Vahedi2016,Vahedi2020}. 
We perform numerical exact diagonalization on Hamiltonian in Eq. (\ref{H}). Average entanglement entropy is then evaluated for $L=140$ sites,   a filling factor $\frac{N}{L}=0.1$ and open boundary conditions  for various values of $\alpha$, in the middle of the energy spectrum.
Fig. \ref{ee} shows the average entanglement entropy in the center of the energy spectrum  or (equivalently for $T\rightarrow
\infty$), for $\alpha$ ranging between $0.2$ and $3$.
The average EE in the middle of the energy spectrum displays a power-law divergence for $\alpha<1$,  $S(l)\sim
l^{\beta}$ with $0<\beta<1$, as can be observed in Fig. \ref{EDsa}, in agreement with the results obtained by  RSRG-X, reported in the previous section.
For $\alpha>1$ the  EE scaling shows a logarithmic enhancement similar to what was found for the ground state, in agreement with RSRG-X. The entanglement scaling exponents $\beta$ obtained by  fitting the 
ED  data are in good agreement with what was found through the RSRG-X procedure especially for lower values of $\alpha$, as seen in Fig. \ref{EDRGcom}.

\section{Entanglement Contour} \label{EC}
A related  interesting quantity  is the contour for the entanglement entropy\cite{Chen2014,eBuruaga2020}. In a lattice,  where a spatial cut separating the chain in two subsystems $A$ and $B$ has been introduced, it is given by a  function $s_A(i)$ which provides information about the contribution of
the $i^{th}$ site in $A$ to the entanglement between $A$ and $B$. By construction, the minimal properties that the contour function must satisfy are:
\begin{equation}
    S_A= \sum_{i \in A} s_A(i) , \hspace*{0.5cm} \text{$s_A(i)>0$},
    \label{contour}
\end{equation}
where the first condition  in Eq. (\ref{contour}) is a normalization, while the second ensures that the contribution of
each site to the entanglement entropy is positive.

Within the RSRG-X framework for excited states of random spin chains, a natural definition of the entanglement contour arises. Namely, the value of $s_A(i)$ on a given site $i$ in $A$ is given by $\ln 2$ if there is a bond  starting at site $i$ and ending in $B$ and if this link is a singlet or an entangled triplet, while  $s_A(i)$ is zero otherwise.

Results for the contour function $s_A(n)$ in the LR random XX chain as a function of the position index $n$ in $A$, as obtained via RSRG-X at infinite effective temperature,
$T \rightarrow \infty$ are displayed in Fig. \ref{contour} for different values of $\alpha$.
Here, we have taken the block $A$ to be the left half of the chain that contains $N/2$ spins. $n$ is the index distance measured from the center of the chain.
The figure shows   $\langle s_{N/2}(n)\rangle $, that is the average contribution of the $n^{th}$ spin to the entanglement entropy between the left half and the right half of the chain.

From Fig. \ref{contour} we see that a strong $\alpha$ dependence appears. Indeed, for $\alpha<1$, $ s_{N/2}\sim n^{-\gamma}$ with $0<\gamma<1$. While for $\alpha>1$ we obtain that $s_{N/2}\sim n^{-1}$.
Since the EE of a subsystem of $n$ spins is given by $S_n=\sum_{i=1}^{n} s_A(i)$, we find,  in the limit of $ N \gg 1$, that for $\alpha>1$ the entanglement entropy diverges logarithmically $S_n \sim \ln(n)$, as it was obtained above via ED and RSRG-X.
Whereas for $\alpha<1$, this yields a power-law growth of EE as a leading term,  $S_n \sim n^{1-\gamma}$.
This result is consistent with $S_n\sim n^{\beta}$ previously found using RSRG-X and ED for $\alpha<1$, since  $\beta\sim 1-\gamma$.
Strikingly, we note that for $\alpha \ll 1$, the entanglement contour  exponent $\gamma$ approaches the bare coupling exponent $\alpha$. Although we do not have an analytic understanding of this behavior, we conjecture that  $\gamma \sim \alpha$ for $\alpha\ll1$.

 \begin{figure}
\centering
\includegraphics[width=.5\textwidth]{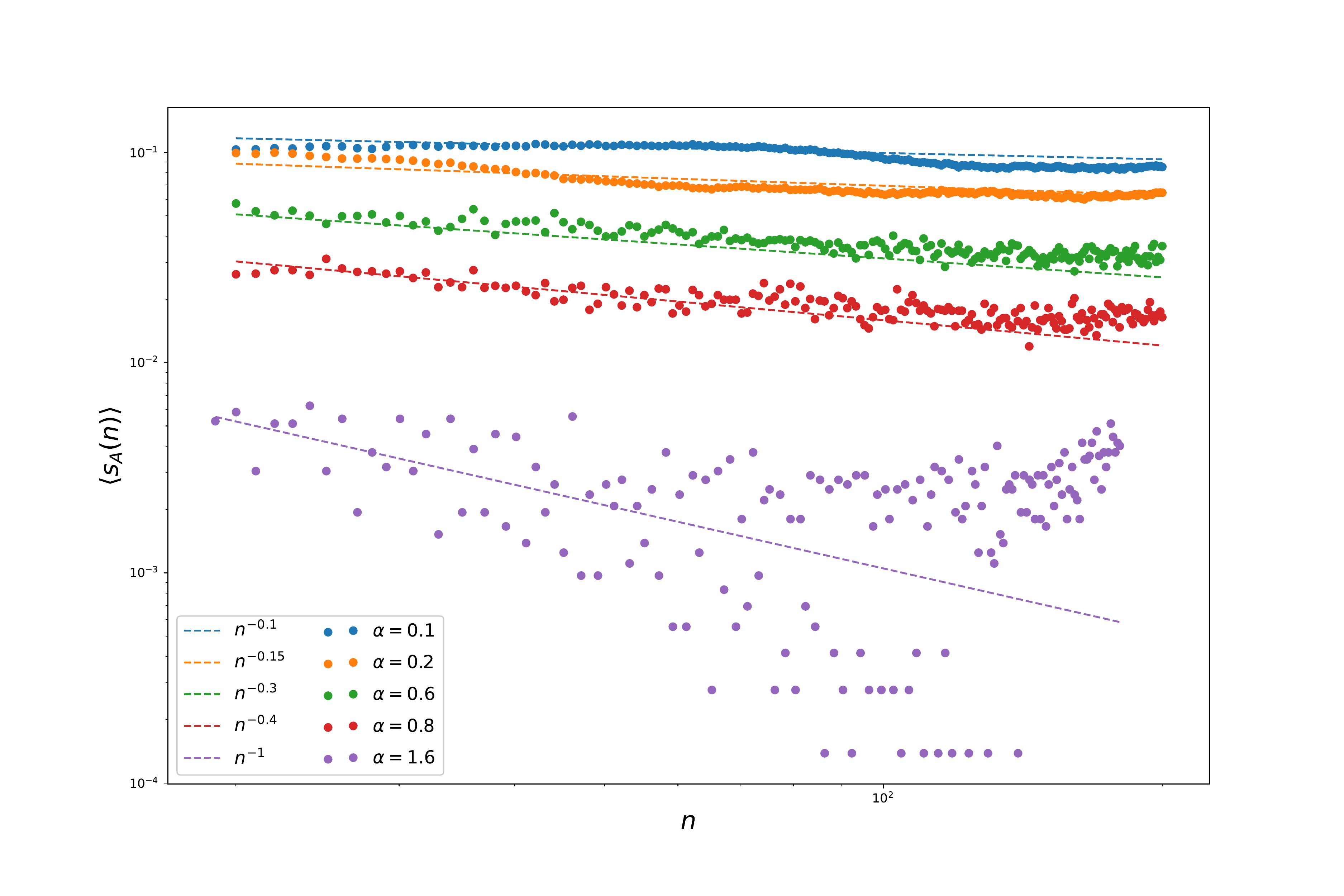}
 \caption{Entanglement contour as obtained via RSRG-X at  infinite effective temperature  for the LR random XX chain for different values of $\alpha$. Data points were obtained for $N=500$ spins randomly placed among $L=5000$ sites. The figure shows the entanglement contour $s_A(n)$ as a function of the position $n$ inside block $A$ of the chain. Here the block $A$ was chosen to be the left half of the chain containing $N/2$  spins. Each data point was obtained by
averaging over $2000$ disorder realizations and $50$ sampled states at each disorder realization.}
\label{eecont}
\end{figure}

\section{Towards a Scaling Theory of  the  Strong  Violation of the Area Law of the Entanglement Entropy}
\label{ST}

 In the previous sections we 
 found  a strong violation of the area law, 
 an anomalous scaling  law for the average entanglement entropy  given by 
 $S_l \sim l^{\beta}$,
 with $\beta (\alpha) < 1$ a decreasing function of $\alpha$ for $\alpha <1$
 both with the  exact diagonalization and with the RSRG-X method.  Both methods 
yield good quantitative agreement for $\beta (\alpha) \sim 1- \alpha$ as shown in 
Fig. \ref{EDRGcom}. Still, it is desirable to get an intuitive understanding 
 why the scaling has this anomalous behavior for $\alpha <1$. To this end
  we aim to formulate an entanglement scaling theory in this section. 
  
In Ref. \onlinecite{Vitagliano2010}  a strong disorder renormalization group procedure was used   to study the ground state entanglement properties of  XX spin chains with
particular local inhomogeneities. There, the couplings were defined such that RG produces
for the ground state a product of concentric singlets, resembling {\it rainbows},   resulting in a volume law scaling of the entanglement entropy. This is now known as the rainbow chain\cite{Ramrez2014,RodrguezLaguna2017}. More recently, disorder was introduced in such a model, resulting in a coexistence of rainbows and dimer bonds  of neighoured spins   for the ground state and giving rise to a weaker area law violation\cite{Alba2019} in form of a power-law $S_l\sim l^{0.5}$. 
This motivates us to  use ideas developed in these papers, and adapt them to the  RSRG-X scheme in order to understand the peculiar algebraic scaling and its dependence on 
$\alpha$ of EE in presence of long-range interactions. 

Let us start from the   set of 
RSRG-X rules Eq. (\ref{updown}), Eq. (\ref{upup}) and Eq. (\ref{jeff}), which we  derived for the Hamiltonian Eq. (\ref{H})  
in order to qualitatively derive the entanglement entropy scaling with subsystem size $l$. 
First, note that the perturbative corrections   for the triplet states  Eq. (\ref{updown}) and Eq. (\ref{upup}) 
are typically larger than 
the ones for singlet states, Eq. (\ref{jeff}). This is especially  true for small values of $\alpha \ll 1$, resulting in larger 
renormalized values of  the couplings $J_{ij}$.

We show now that the RG rule for the {\it entangled triplet state}
Eq. (\ref{updown}) favours the formation of rainbows. Here,  a rainbow is  
defined to be any pair of spins with a bond which is not connecting neighboring sites and is thus not a dimer (bond connecting
neighboring sites). To illustrate this, assume that a pair $(n, n+1)$ is projected onto the entangled triplet state $\frac{1}{\sqrt{2}}(|\uparrow\downarrow\rangle+ |\downarrow\uparrow\rangle)$.
Then, based on the observation that the perturbative corrections are typically smaller for  sites farther away from each other,  the coupling $J_{n-1,n+2}$ 
is likely to take the largest correction. This  leads to the fact that the next decimated bond would e between the sites $(n-1, n+2)$. 
The RG-rule for the nonentangled triplet states Eq. (\ref{upup}) also tends to create long range bonds, since  couplings next to the decimated bond are diminished  more than the farther ones.

Thus, the RG rules  lead to a state  of consecutive dimers   which are spanned over by   rainbows (bonds connecting farther sites). This picture is expected to  become more  accurate 
 with smaller $\alpha$,  since the renormalization  corrections increase  
 with smaller $\alpha$, thereby 
 favoring more frequent rainbow formation.
 
\begin{figure}
\centering
\includegraphics[width=.45\textwidth]{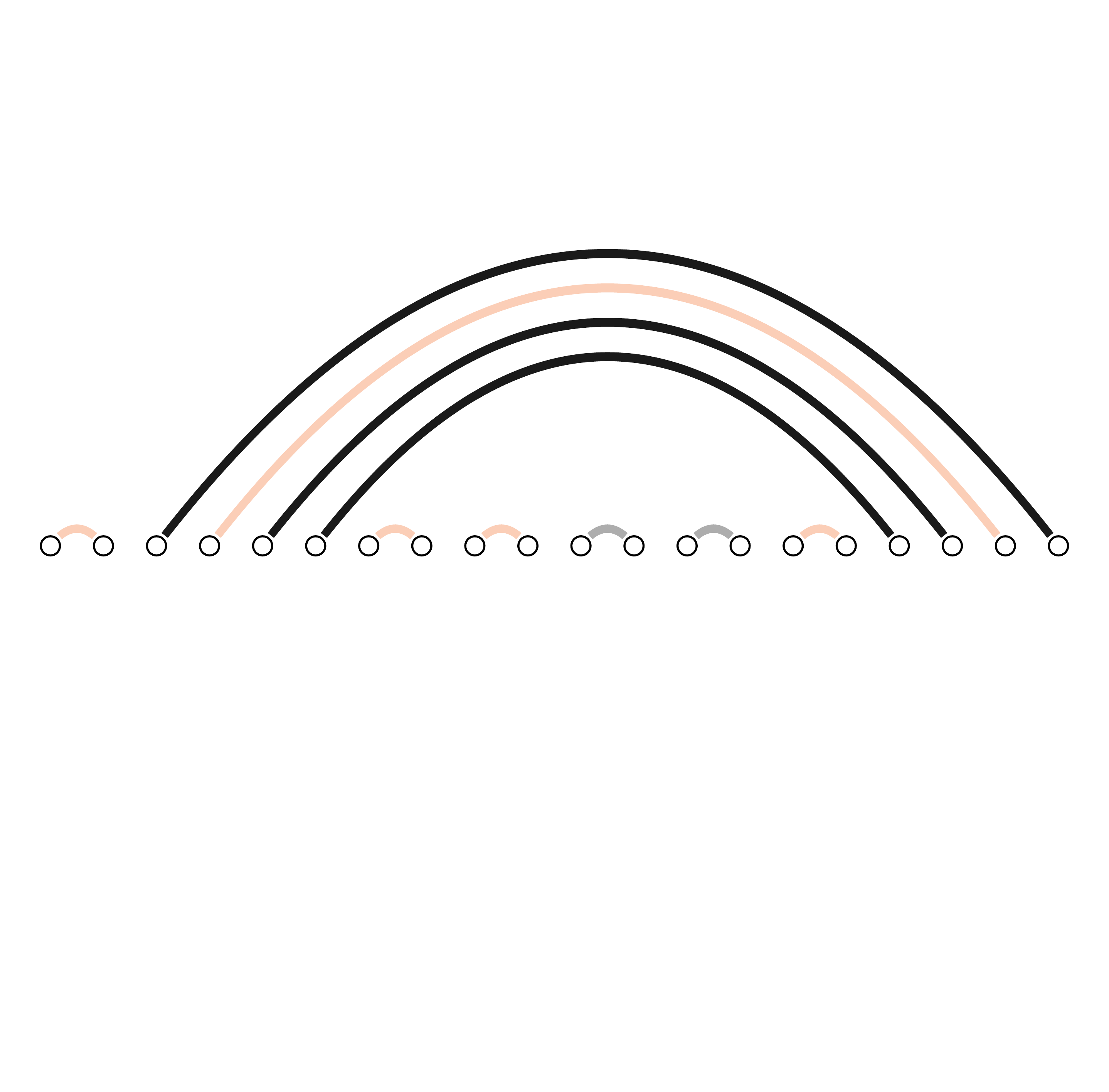}
\vspace*{-3.8cm}
\caption{Illustration of a realization at $N=20$. The orange bonds represent pairs projected onto
 $ |\uparrow\uparrow\rangle)$ or $|\downarrow\downarrow\rangle)$, while black links stand for $\frac{1}{\sqrt{2}}(|\uparrow\downarrow\rangle+ |\downarrow\uparrow\rangle)$ and the grey ones are singlets }
\label{figexp}
\end{figure}

In order to evaluate the entanglement scaling  with subsystem sizes, we define $m_d$ as the number of  consecutive  dimers regardless of the state they are projected onto.
A configuration for $N=20$ is shown in Fig. \ref{figexp}.
In this configuration, there are two dimer   regions. The first one is made of one bond $m_d=1$ (the bond connecting the sites $n=1$ and $n=2$), while the other dimer region is composed of $5$ bonds (from spin at site $n=7$ to $n=16$) so that $m_d=5$. While there is one rainbow region with $m_r=4$. 
Different colors indicate to which state the pair is projected to. 

\begin{figure}

\includegraphics[scale=0.33]{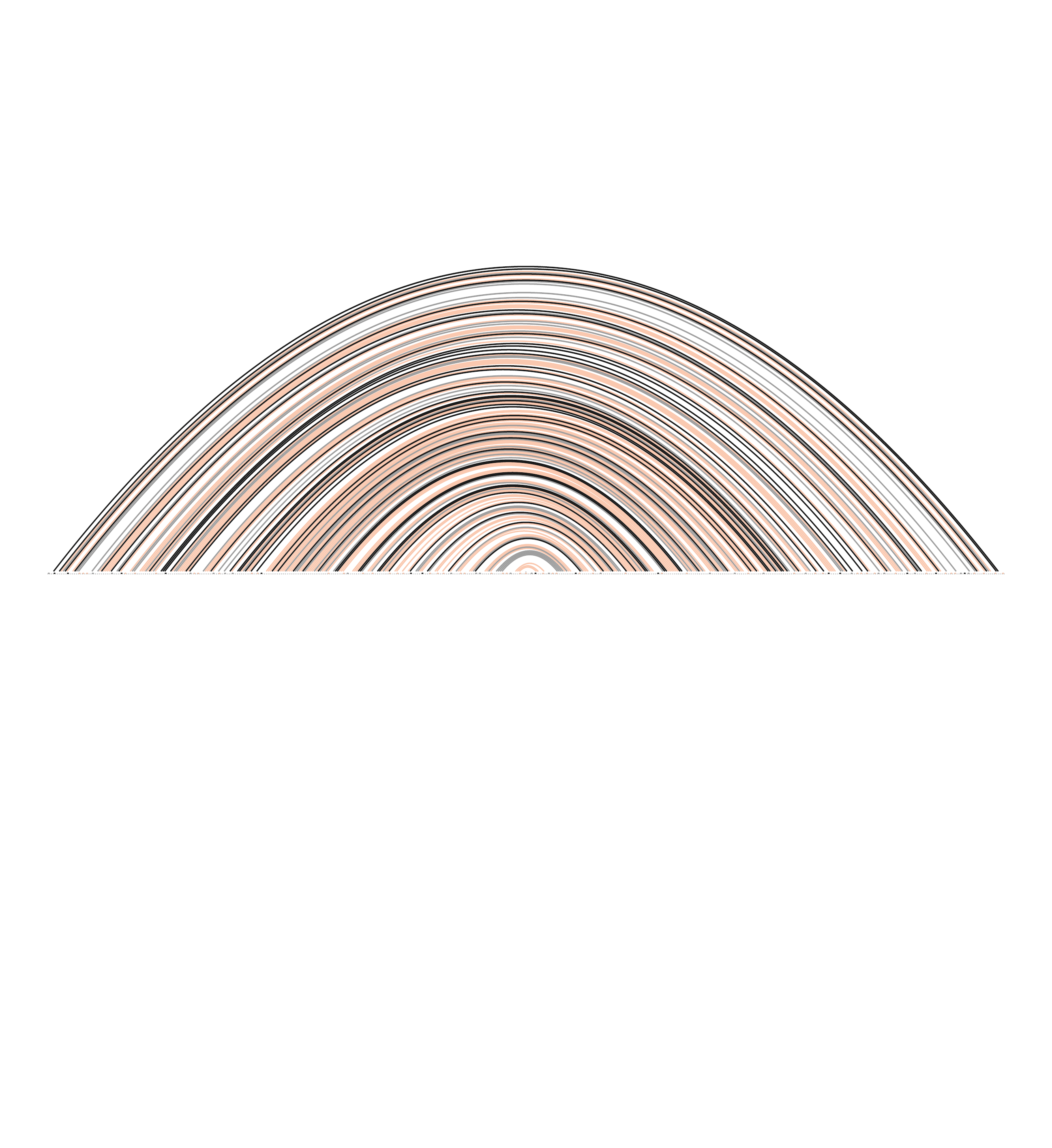}
\vspace*{-5.1cm}

\includegraphics[scale=0.4]{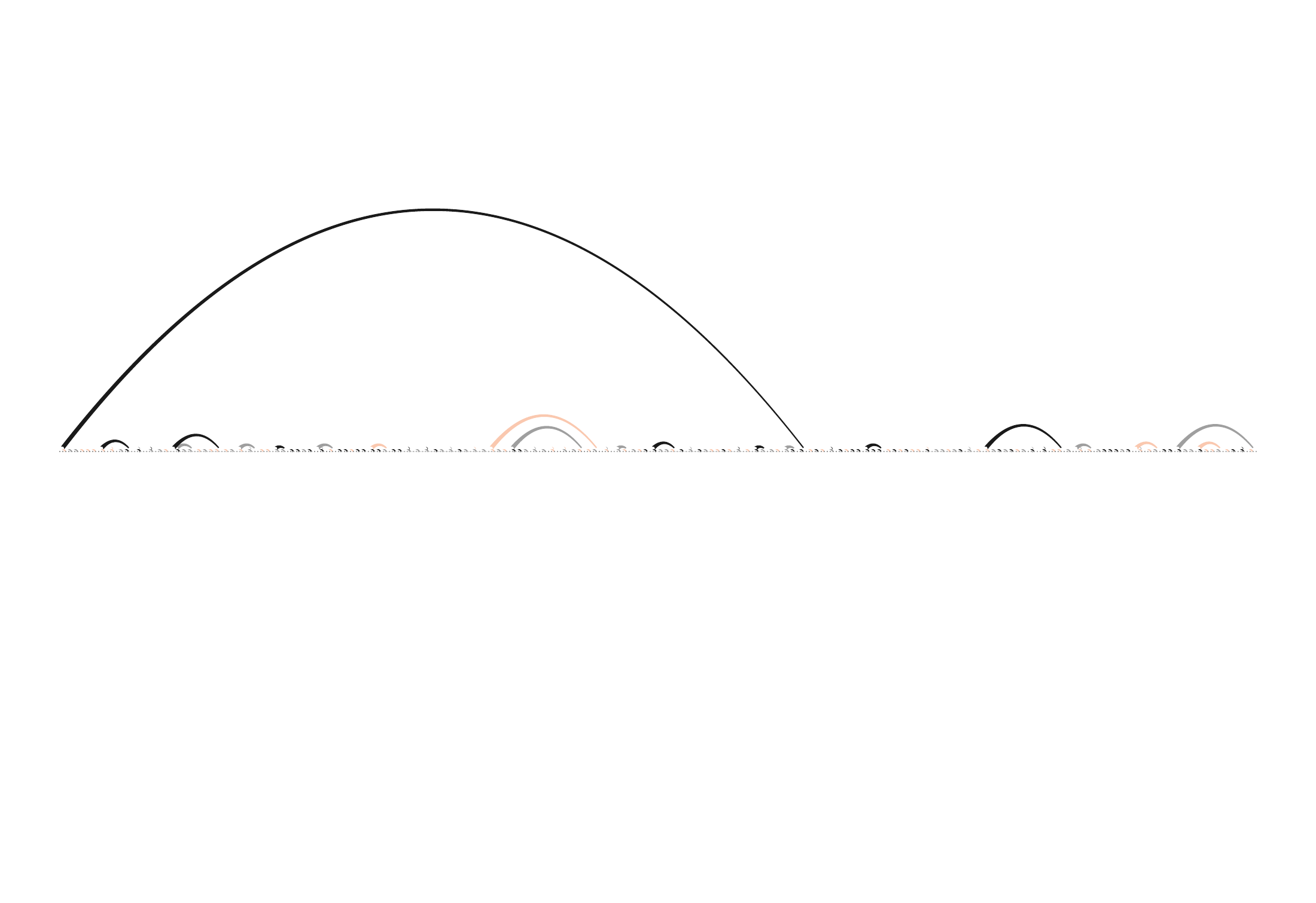}
\vspace*{-3cm}
\caption{Top: Typical eigenstate produced by RSRG-X for $\alpha=0.1$ obtained  for $N=400$ spins and a filling factor $\frac{N}{L}=0.1$. The bonds in black are pairs projected onto $\frac{1}{\sqrt{2}}(|\uparrow\downarrow\rangle+ |\downarrow\uparrow\rangle)$  while the orange links are in the states $ |\downarrow\downarrow\rangle)$ or $|\downarrow\downarrow\rangle)$, and the grey ones are singlets.
Bottom : Same but for $\alpha=1.6$}
\label{realconfiga}
\end{figure}

A finite   entanglement entropy is due to  rainbow links which are ranging across the boundary  of the subsystem and which are projected either onto the singlet state or the entangled triplet state (in the large system limit we may disregard the entanglement from a single dimer which may cross the boundary).
We assume that the rainbows are connecting sites symmetric around the middle of the chain, and are therefore links connecting sites $i, N-i$ as can be qualitatively observed in Fig. \ref{realconfiga} (top), for sufficiently small $\alpha$.

Let the spins in a subsystem be numbered from $1$ to $n$ with $n<N/2$ and consider the site $i$ and its bond. As dimers do not contribute to the EE, one can  write
\begin{equation}
    \langle S_n \rangle\sim \frac{1}{2}\ln2 \sum_{i=1}^{n} P(\text{$i^{th}$ bond is a rainbow}), 
\end{equation}
where the factor $1/2$ is due to the fact that only two states out of the $4$ possible contribute to EE, and $\ln 2$ is the amount of entanglement generated by a singlet or an entangled triplet. Note that here, we use $n$ as the index distance.
Assuming that the distribution of rainbows along the chain is homogeneous,
\begin{equation}
    \langle S_n \rangle\sim \frac{1}{2} (\ln 2) n P(\text{$i^{th}$ bond is a rainbow}). 
\end{equation}
 Thereby, as a link belongs  either to a rainbow region or to a dimer region, 
 $P(\text{a link is a rainbow})\sim \frac{n_r }{n_r+n_d} $.
 Where $n_r$ is the number of rainbow bonds and $n_r+n_d$ is the total number of bonds within a subsystem of $n$ spins.
If we further assume that the rainbows (and the dimer) regions  are homogeneously distributed, $n_r= \langle m_r \rangle N_r$ rainbow bonds and $n_d= \langle m_d \rangle N_r$, where $N_r$ is the number of rainbow (dimer) regions. 
 We thus obtain that  the EE of a subsystem of $n$ spins  can  be approximated as
\begin{equation}
    \langle S_n \rangle\sim \frac{1}{2} (\ln2) n \frac{\langle m_r \rangle }{\langle m_r \rangle+\langle m_d \rangle}. 
    \label{EErbbubble}
\end{equation}
In the case of the rainbow chain\cite{Vitagliano2010}, one has $\langle m_r \rangle=n $ and $\langle m_d \rangle=1$, yielding a volume-law entanglement $S_n\sim n$.   
Thus, it  remains  to evaluate $\langle m_d \rangle$ and $\langle m_r \rangle$. 
 
 Let us first derive    $Pr(m_r=k)$, the probability mass function (PMF) of the rainbow regions length. We assume that rainbow regions are formed successively along the RSRG-X flow. Since a newly formed bond can only be a rainbow or a dimer, and the $3$ possible triplet projections favour rainbow region expansion, the probability that a region is formed of $k$ rainbows is equal to the probability of forming successively $k$ rainbows with probability $3/4$ for each, then forming a dimer   with probability $1/4$.
The PMF $Pr(m_r=k)$ can thus be approximated by a geometric distribution, with $p=1/4$, yielding $ Pr(m_r=k) = \left(3/4\right)^k  (1/4)$.

\begin{figure}
\includegraphics[width=.5\textwidth]{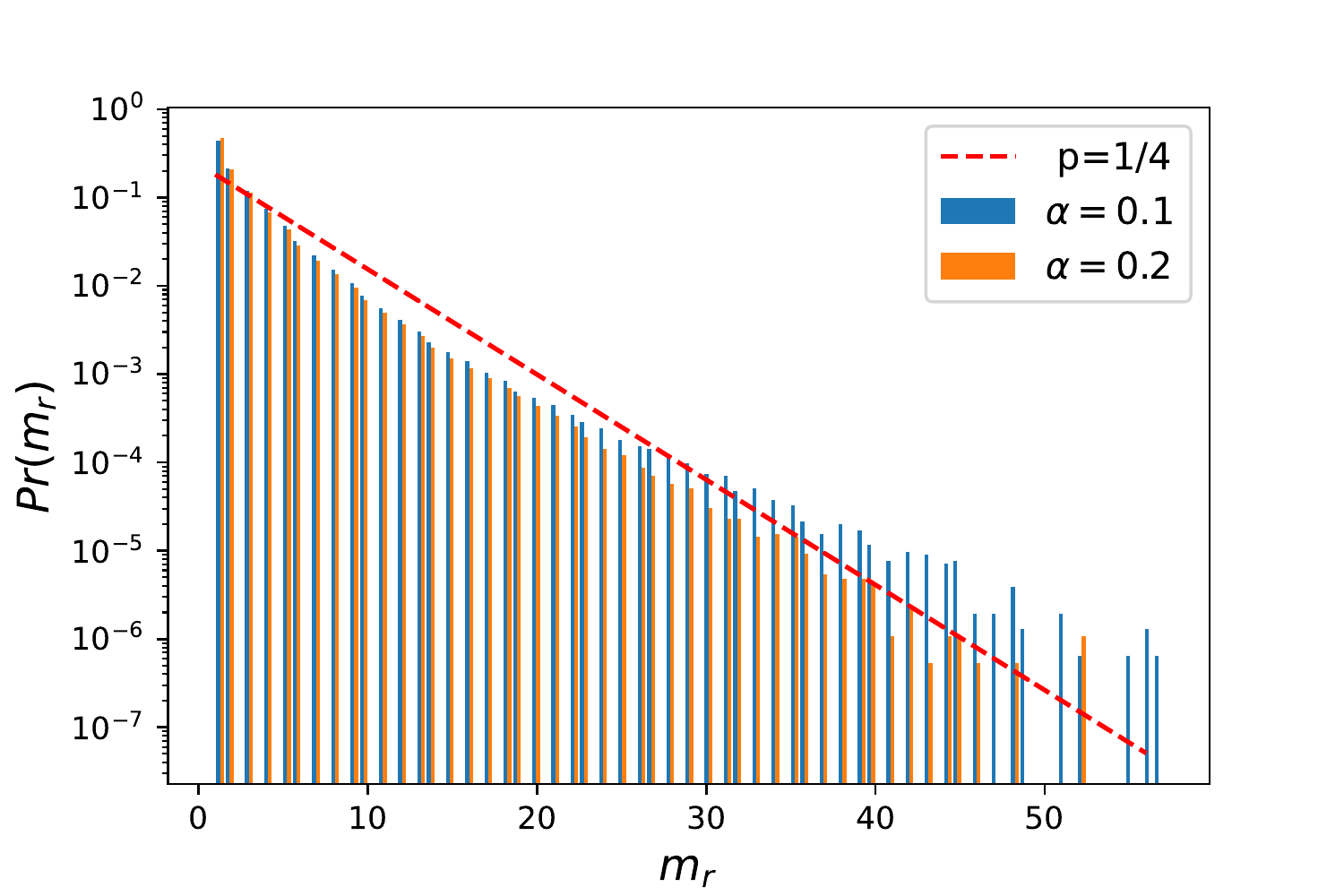}
\includegraphics[width=.5\textwidth]{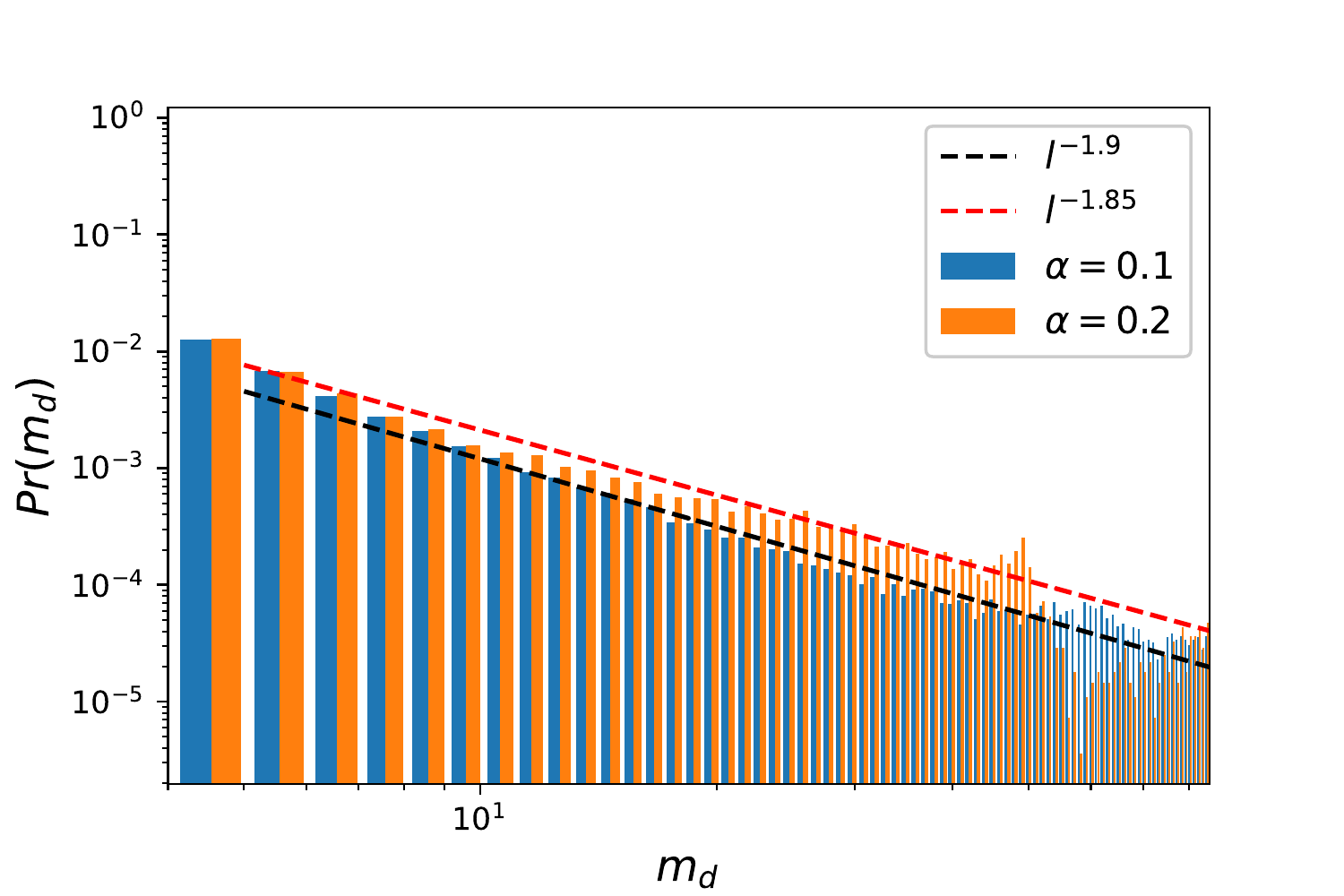}

 \caption {Top:Probability distribution of the lengths of the rainbow regions $P(m_r)$ for $\alpha=0.2$ and $\alpha=0.1$. The red dashed line is a geometric distribution with $p=1/4$.
 Bottom: Probability distribution of the length of the dimer regions $P(m_d)$. The histograms were obtained for $30000$ realizations, $N=300$ and a filling factor $\frac{N}{L}=0.1$}
\label{Histlblr}
\end{figure}

Fig. \ref{Histlblr} (top) shows the decay of $P(m_r)$ for $\alpha=0.1$ and $\alpha=0.2$ as obtained via RSRG-X at infinite effective temperature. Here, a logarithmic scale was used for the y-axis. Clearly, the geometric distribution with $p=1/4$ provides a good approximation for $P(m_r)$ for both $\alpha=0.1$ and $\alpha=0.2$.
Thus, as $P(m_r)$ decays very quickly (exponential decay in the continuum limit) $\langle m_r \rangle =\sum_{k=0}^{n/2}  k P(m_r=k)  $   is a constant not dependent on $n$ for sufficiently large $n$.

Moreover, $\langle m_d\rangle$ can be directly related to the entanglement contour $s_{N/2}(n)$. Indeed, as the dimer bonds do not contribute to EE, $s_{N/2}(n)$ is the probability that the link starting at the $n^{th}$ spin is a rainbow, which was obtained above under the homogeneity assumption  $s_{N/2}(n)\sim \frac{\langle m_r \rangle }{\langle m_r \rangle+\langle m_d \rangle} $.
As $\langle m_r \rangle\sim 1$, and since for $\alpha \ll 1$, $s_{N/2}(n)\sim n^{-\alpha}$, we obtain that for $\alpha=0.1$, $\langle m_d \rangle \sim \frac{1}{s_{N/2}(n)}\sim n^{0.1}$, 
while for $\alpha=0.2$, this implies $\langle m_d \rangle \sim n^{0.2}$. Note that for the rainbow chain this yields $s_{N/2}(n)\sim \frac{\langle m_r \rangle }{\langle m_r \rangle+\langle m_d \rangle}\sim 1$ which is consistent since all the sites equally contribute to EE.

To confirm this result we implement the RSRG-X procedure at infinite effective temperature and numerically compute the histograms of the lengths of dimer  regions for sufficiently small values of  $\alpha$. Results are shown in Fig. \ref{Histlblr} for $\alpha=0.1$ and $\alpha=0.2$.
From Fig. \ref{Histlblr} (bottom) we obtain that $P(m_d)\sim m_d^{-1.9}$ for $\alpha=0.1$. Leading to 
$\langle m_d \rangle = \sum_{k=1} ^{n}  k P(m_d=k)  \sim n^{0.1}$, as obtained using the entanglement contour, therefore
\begin{equation}
    \langle S_n \rangle \sim  \frac{n}{\langle m_d \rangle}\sim n^{0.9}.
\end{equation} 
Similarly Fig. \ref{Histlblr} indicates that for $\alpha=0.2$, $P(m_d) \sim m_d^{-1.85}$, which thereby leads to  $ \langle S_n \rangle \sim n^{0.85}$. 
The power exponents $\beta$ obtained through this description  therefore  turn out to be in good agreement with those we have obtained for $\alpha\ll 1$, as can be seen in Fig. \ref{EDRGcom}.   

We emphasize that this picture is expected to break down for higher values of $\alpha$, $0.4<\alpha<1$ as the corrections to the couplings are less significant, leading to a less probable rainbow formation, and therefore lower entanglement.
Whereas for $\alpha>1$ the perturbative corrections are  small enough to flow to a strong disorder  fixed point of random bonds (similar to  the ground state, which is a random singlet state). Since each bond may be with probability $1/2$ an entangled state, this 
explains the logarithmic enhancement of EE.
Fig. \ref{realconfiga} shows two typical eigenstate structure as obtained by RSRG-X for $N=400$ spins, for a small value  $\alpha=0.1$ and a higher value $\alpha=1.6$. For $\alpha=0.1$, we clearly observe a proliferation of rainbow-bonds, inducing links connecting farther sites in the chain, and therefore generating  stronger entanglement $S_n\sim n^{0.9}$, while for $\alpha=1.6$ dimer-bonds are dominant, and spins which are paired with  far away spins are very rare, similar to the $T=0$ random singlet phase, yielding a logarithmic enhancement of EE.


\section{Energy and Effective Temperature dependence of Entanglement Entropy } \label{ETEE}

As we have so far only considered the model in the middle of the many-body spectrum, we now explore the half-chain entanglement entropy $S_{\epsilon}(L/2)$ of the eigenstates at different energy densities, aiming to evaluate how  the entanglement  depends on the energy scale.  

{\it Energy Dependence.}
At first, we  employ exact numerical diagonalization, we sample $50$ states close to a target energy $\epsilon$ for every disorder realization, and 
then calculate the average entanglement entropy in the middle of the chain $l= L/2$.
The energies $\epsilon$ are normalized to be in the interval $[0,1]$ as $\epsilon=(E-E_{min} )/(E_{max} - E_{ min} )$,
where $E_{min}$ is the ground state energy.
\begin{figure}
\includegraphics[width=.5\textwidth]{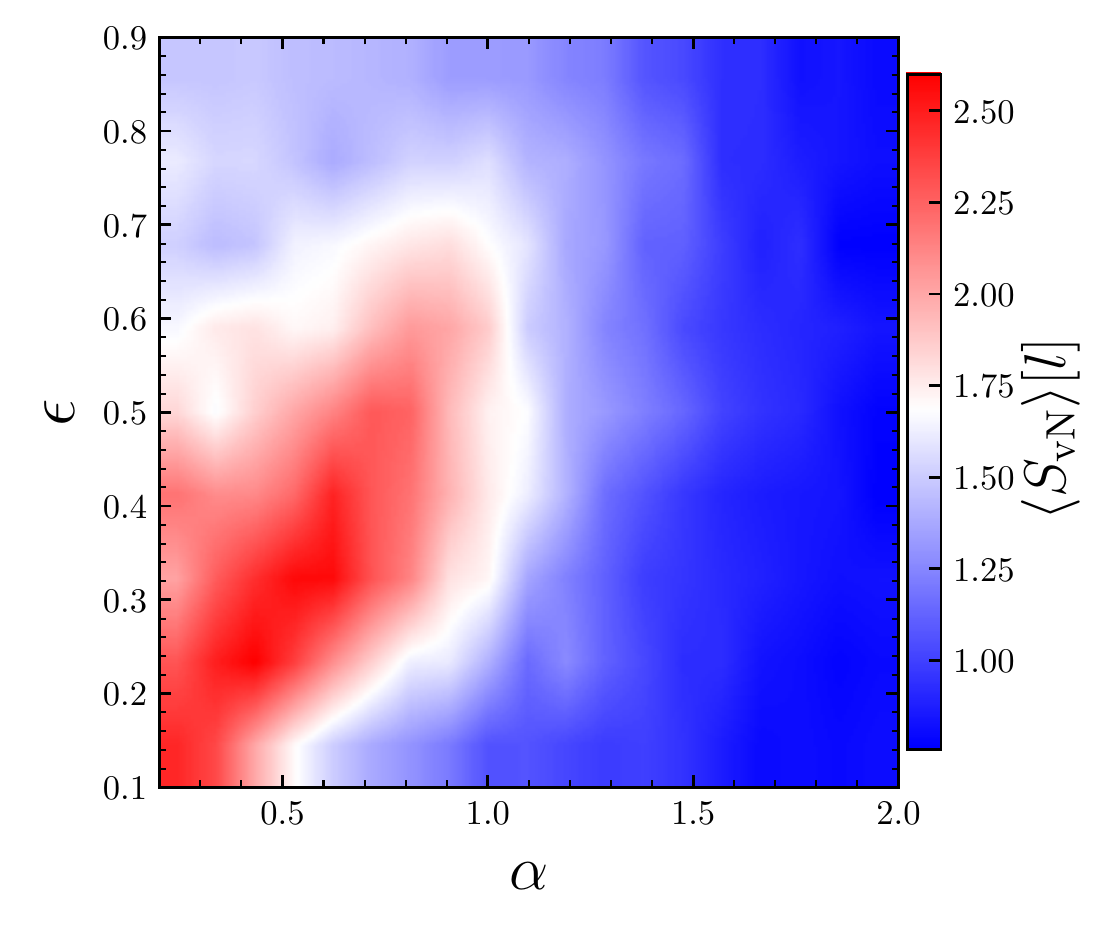}
\caption{ Half-chain entanglement entropy $S_{l/2}$ for different values of $\alpha$ and at different energy scales $\epsilon$. The results are obtained for $N=18$, a filling factor $\frac{N}{L}=0.1$ and $M=1000$ disorder realizations}
\label{EEenergy}
\end{figure}
Results  For $\epsilon \in [0.1,0.9]$, $N=18$ spins and $L=180$  are shown in Fig. \ref{EEenergy}.
First, we note that around the middle of the energy spectrum  $\epsilon \in [0.3,0.6]$, a crossover between a regime with higher entanglement and a less entangled phase occurs at $\alpha^{*} \approx 1$,  confirming the previously obtained results via both ED and RSRG-X. 
However, for smaller energies, we observe that there is a shift towards smaller
values of $\alpha^{*} $. Namely, for $\epsilon \in [0.1,0.3]$ we see that $\alpha^{*} $
ranges between $0.6$ and $0.8$, while the crossover is not  observed in the half-chain EE at very high energies.

\begin{figure}
\includegraphics[width=.5\textwidth]{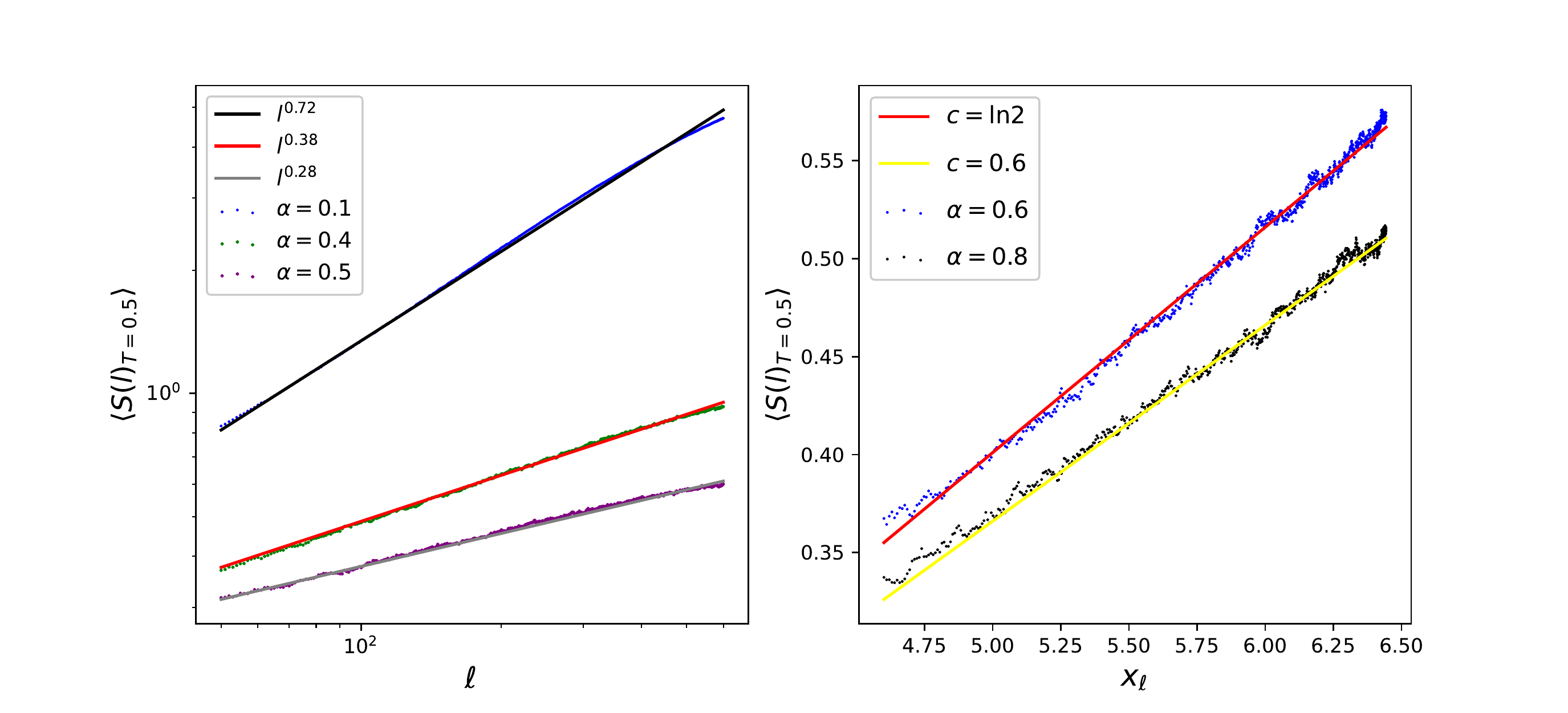}
\caption{ Average entanglement entropy of excited eigenstates at finite effective temperature $T=0.5$  as a function of the partition length $l$ (physical distance), obtained via RSRG-X for $N=200$ spins. The filling factor $\frac{N}{L}=0.1$ was considered for various values of $\alpha$. 
The average was evaluated over $ 5000$ disorder realizations, and $100$ sampled states for each disorder realization. }
\label{EEltemp}
\end{figure}

{\it Effective Temperature Dependence.}
Next, let us implement  RSRG-X  to explore the effective temperature dependence of  EE. To this end,  we employ RSRG-X  as  described in section \ref{RSRGX}, but this time at finite effective temperature
parameter $T$. We fix  $T=0.5$, and sample $100$ eigenstates for each disorder realization via the RSRG-X procedure, then calculate the resulting  EE as a function of subsystem sizes $l$ for different values of $\alpha$.
Results are shown in Fig. \ref{EEltemp}. For $\alpha<0.6$ we observe a power law growth of EE  $S_l\sim l^{\beta}$ as a function of the partition length. The exponents $\beta$ are however lowered as compared to the $T \sim \infty $ case, as can be seen in Fig. \ref{EEltemp} (left) where average EE is plotted against subsystem  sizes, and turns out to be linear on the log-log scale. While,
for $\alpha\geq 0.6$ EE displays a logarithmic enhancement as can be seen in Fig. \ref{EEltemp} (right) where EE is plotted as function of the logarithm of the chord distance $x_l=\ln(\frac{L}{\pi} sin(\frac{\pi l }{L}))$ for $\alpha=0.6$ and $\alpha=0.8$
resulting in a Cardy-law with central charges  $c_{eff}=\ln 2$ and $c_{eff}=0.6$ respectively.

These results indicate that at lower effective temperature the highly entangled phase survives only up to $\alpha^{*}\approx 0.6$.
This can be explained by the rarefaction of projections onto $\frac{1}{\sqrt{2}}(|\uparrow\downarrow\rangle+ |\downarrow\uparrow\rangle)$ for smaller effective temperatures, therefore inducing a less prevalent rainbow formation, and thus, a smaller entanglement entropy  for a given $\alpha$. 

\section{Conclusion} \label{C}
We  introduced  a real space renormalization group scheme for excited eigenstates, a modified 
RSRG-X, for   XX spin chains with random interactions decaying as a power-law with distance.
The entanglement entropy (EE) is calculated using RSRG-X and exact diagonalization for different values of the power exponent $\alpha$  in the middle of the many-body spectrum. 
The results obtained via RSRG-X and ED are in good agreement and show that 
  the average excited eigenstate entanglement entropy  grows as a power-law with the subsystem size
  $l$, $S(l)\sim l^{\beta}$, with power  $0<\beta<1$ a decreasing function of $\alpha$ for $\alpha<\alpha^*$, while a logarithmic enhancement of EE entropy is observed for $\alpha>\alpha^*$. 
  We find that in the middle of the many-body spectrum $\alpha^* \approx 1$ coincides with the delocalization transition $\alpha_c$, which we  derived from  the level spacing statistics 
  obtained with ED for system sizes up to $N\sim 18$ spins to be  at 
$\alpha_c \approx 1$ in the middle of the spectrum. 

Using RSRG-X, we also investigate the entanglement contour for different values of $\alpha$
and find it to decay as a power of the subsystem size $l$ with power $\gamma$. 
We find good agreement for the conjecture that 
$\gamma \approx \alpha$. In an effort to derive this conjecture, we  
 suggest a   scaling theory based on the RG-rules structure for $\alpha \ll 1$. 
 To illustrate and 
 support this approach, we  show  typical eigenstate configurations as obtained with RSRG-X for $\alpha=0.1$ and $\alpha=1.6$. 
 
 In addition, we  compute the half-chain EE at energy densities, which 
corresponds to a position away  from the middle of the many body spectrum, 
and implement RSRG-X  at lower effective temperature  $T=0.5$, corresponding to smaller energy density. The results indicate a crossover  between a phase with strongly enhanced entanglement for $\alpha < \alpha^* \approx 0.6$  and a regime with logarithmic scaling of EE for $\alpha > \alpha^* \approx 0.6$. Thus, 
 we find indications that $\alpha^* \le \alpha_c$. 
 This observation is consistent with the fact that in Ref. \cite{Mohdeb2020} we found 
  in the ground state of the LR AFM coupled disordered spin chain that the area law violation of EE is 
   logarithmic for all  $\alpha$, so that $\alpha^* \rightarrow 0,$ while we had found previously that the delocalization transition in its ground state occurs at $\alpha_c \approx 1$\cite{ours,ourPRB}.
In summary, these  results indicate that  for   $ \alpha>\alpha_c$, the model behaves 
 similarly as the  nearest neighbor random bond  model \cite{Huang2014}, which   
was found to  be in  a quantum critical glass phase\cite{Vasseur2015}, a regime where arbitrarily high energy excited states exhibit  power-law decaying correlation functions and logarithmic divergence in entanglement entropy.

As excited eigenstates are states that participate in the dynamics of the system, understanding their properties is crucial in order to characterize quantum phase transitions. Therefore, 
 building on these results for the   higher energy eigenstates, 
 we can aim in future research  to study quantum quench dynamics 
 in random spin chains with power law long range couplings.

\section{Acknowledgement}           
We  gratefully acknowledge  the support from Deutsche Forschungsgemeinschaft (DFG) KE-807/22-1.

\newpage

\bibliography{ref}

\end{document}